%% file: kappa.tex
\documentclass[a4paper,12pt]{article}
\pdfoutput=1 

\usepackage{jheppub} 

\usepackage[english]{babel}
\usepackage[utf8]{inputenc}
\usepackage[T1]{fontenc}
\usepackage{type1cm}
\usepackage{type1ec}
\usepackage{amsfonts}
\usepackage{nicefrac}

\usepackage{pgf}
\usepackage{tikz}
\usepackage{scalefnt}
\usepackage{caption}
\usepackage{gnuplot-lua-tikz}

\newcommand {\dd}{\mathrm{d}}

\newcommand {\ii}{\mathrm{i}}
\newcommand {\RE}{\operatorname{Re}}

\newcommand {\Tr}{\operatorname{Tr}}

\newcommand{\sfreqq}[1]{\sum\hspace{-0,55cm}\int\limits_{#1}}
\newcommand{\abs}[1]{\lvert {#1}\,\rvert} 

\newcommand{\e}{\mathrm{e}}
\newcommand{\colourN}{N_{\mathrm{c}}}
\newcommand{\latq}{\widetilde{q}}
\newcommand{\latk}{\widetilde{k}}

\title{\boldmath The second order hydrodynamic
transport coefficient $\kappa$ for the gluon plasma from the lattice}
\author{Owe Philipsen,
Christian Sch\"afer}
\affiliation{Institut f\"ur Theoretische Physik, Goethe-Universit\"at Frankfurt,\\
Max-von-Laue Str. 1, 60438 Frankfurt am Main, Germany}
\emailAdd{philipsen,cschaefer@th.physik.uni-frankfurt.de}

\abstract{The quark gluon plasma produced in heavy ion collisions behaves
like an almost ideal fluid described by viscous hydrodynamics with a number of
transport coefficients. 
The second order coefficient $\kappa$ 
is related to a Euclidean
correlator of the energy-momentum tensor at vanishing frequency and
low momentum. 
This allows for a lattice determination without
maximum entropy methods or modelling, but the required
lattice sizes represent a formidable challenge. We calculate
$\kappa$ in leading order lattice perturbation theory and simulations 
on $120^3\times 6,8$ lattices
with $a<0.1$ fm.
In the temperature range $2T_\mathrm{c}-10T_\mathrm{c}$ we find
$\kappa=0.36(15)T^2$. The error covers both
a suitably rescaled AdS/CFT prediction as well as, remarkably, 
the result of leading order perturbation theory. This suggests that
appropriate noise reduction methods on the lattice and 
NLO perturbative calculations could provide
an accurate QCD prediction in the near future.
}

\begin{document} 
\maketitle
\flushbottom

\section{Introduction}
\label{sec:introduction}

One of the major findings of the experimental heavy ion programme \cite{Adcox:exp1,Back:exp2,Arsene:exp3,
Adams:exp4} is that QCD matter at high temperatures and low densities behaves as a nearly ideal
fluid with very low viscosity. This conclusion is based on the fact that experimental data are
excellently described by relativistic hydrodynamics, with transport coefficients fitted to the data
\cite{Teaney:exp5,Huovinen:exp6,Kolb:exp7,Hirano:exp8,	Kolb:exp9,Heinz:exp10}. Unfortunately,
theoretical predictions of transport coefficients from the fundamental theory QCD remain very
difficult \cite{Meyer:review}. Up to a few times the transition temperature to the quark gluon
plasma, the QCD coupling is not weak enough for perturbative methods to apply, which predict a less
ideal fluid \cite{Baym:weak_coupling,Arnold:weak_coupling}. On the other hand, results in the
opposite strong coupling limit can be obtained by AdS/CFT duality 
methods in certain supersymmetric models
\cite{Kovtun:sym,Policastro:sym}, but these do not correspond to QCD directly.

Unfortunately, lattice simulations of real time quantities are in general severely limited by the
need for analytic continuation. Calculations of spectral functions on the lattice based on maximum
entropy methods \cite{Asakawa:mem,Aarts:sigma} or a model ansatz \cite{Nakamura:ansatz,Meyer:eta,Francis:sigma}
require both functional input and high accuracy data
to sufficiently constrain the results. An exception to this conceptual difficulty are the three
second-order hydrodynamic coefficients $\kappa,\lambda_3,\lambda_4$
\cite{Bhattacharyya:2012nq,Jensen:eucl_TK,Banerjee:eucl_TK,Son:kappa, Romatschke:nonconformal},
which can be related to Euclidean correlation functions through Kubo formulae. They have recently
been computed to leading order in a weak coupling expansion in \cite{Moore:kappa}, where also
possibilities for a lattice determination were discussed. The coefficients $\lambda_3, \lambda_4$
are related to three-point functions, which are still too costly to numerically evaluate. 

Here we present a first  attempt to determine $\kappa$ from the momentum expansion of a suitable
two-point function in a lattice simulation. In order to approach the zero momentum limit, very
large lattices are required, demanding an enormous numerical effort already in pure gauge theory.
While the errors on our result are thus still too large to be satisfactory, our work demonstrates
that the determination of the second order coefficients is possible without conceptual difficulties
and should be improved in the future with appropriate noise reduction methods. 
Interestingly and in
contrast to the first order transport coefficients, 
the lattice result for $\kappa$ is within error bars
compatible with the perturbative weak coupling result. It is also compatible
with a suitably rescaled AdS/CFT result.

In section \ref{sec:continuum} we briefly summarise the relation between the transport coefficient
$\kappa$ and a Euclidean correlator of the energy-momentum tensor, in section \ref{sec:lattice} this
is carried over to the lattice formulation, including a leading order perturbative evaluation and a
discussion of renormalisation. Section \ref{sec:results} contains the numerical results of our
simulations. 

\section{\texorpdfstring{The transport coefficient $\kappa$}{The transport coefficient kappa}}
\label{sec:continuum}
The definition of transport coefficients is based on a gradient expansion of the energy-momentum
tensor in relativistic hydrodynamics, but their respective values have to be determined from
experiment or an underlying theory. In the case of the quark-gluon plasma this underlying theory is
QCD. In this section we review the connection between the transport coefficient $\kappa$ and a
Euclidean correlator in QCD, which allows for a direct computation of $\kappa$ without resort to
maximum entropy methods or functional input for the spectral function.

\subsection{Relativistic hydrodynamics}
The basic quantity in relativistic hydrodynamics is the energy momentum tensor (for a review, see
\cite{Romatschke:hydro}), which can be decomposed into an ideal part $T_0^{\mu\nu}$ and a
dissipative part $\Pi^{\mu\nu}$
\begin{align}
 T^{\mu\nu}=T_{(0)}^{\mu\nu} + \Pi^{\mu\nu}.
\end{align}
The ideal part is determined by the hydrodynamic degrees of freedom, wich are the energy density
$\epsilon$, pressure $p$, the fluid's four velocity $u_\mu$ and the metric tensor $g_{\mu\nu}$.
Lorentz symmetry and the identifications $T_{(0)}^{00}=\epsilon$, $T_{(0)}^{0i}=0$ and
$T_{(0)}^{ij}=p\,\delta^{ij}$ in the local rest frame restrict its form to
\begin{align}
 T_{(0)}^{\mu\nu} = \epsilon u^\mu u^\nu + p\left(g^{\mu\nu}
		    +u^\mu u^\nu\right).
\end{align}
The dissipative contribution consists of a traceless part $\pi^{\mu\nu}$ and a part with
non-vanishing trace $\Pi$
\begin{align}
 \Pi^{\mu\nu} = \pi^{\mu\nu} + \left(g^{\mu\nu} + u^\mu u^\nu \right) \Pi.
\end{align}
The former has been specified for a non-conformal fluid in a second order gradient expansion within
$N=4$ Super-Yang-Mills theory
\cite{Romatschke:nonconformal}
\begin{align}
  \pi^{\mu\nu} = -\eta \sigma^{\mu\nu} + \eta \tau_{\pi}\left( \left< D
  \sigma^{\mu\nu} \right> + \frac{\nabla \cdot u}{3}\sigma^{\mu\nu} \right) +
  \kappa \left( R^{\left<\mu\nu\right>} - 2u_\alpha u_\beta
  R^{\alpha\left<\mu\nu\right>\beta} \right) + \ldots~.
  \label{eq:def_kappa}
\end{align}
Besides the shear viscosity $\eta$ and the relaxation time $\tau_\pi$,  to second order the
transport coefficient $\kappa$ enters the expansion and couples to the symmetrized Riemann curvature
tensor $R$, its contractions and the fluid's four velocity $u^\mu$. For explanations of $\nabla$,
$\sigma^{\mu\nu}$, $D$ and further terms in the expansion we refer to
\cite{Romatschke:nonconformal}. Note that even in flat spacetime the transport coefficient $\kappa$
has a non-vanishing value \cite{Romatschke:sumrules,Moore:kappa}. 

\subsection{Thermal field theory}
\label{subsec:kappaQCD}
For the computation of the transport coefficient $\kappa$ from QCD a relation between its definition
in relativistic hydrodynamics and thermal field theory is necessary. This can be achieved by
considering the fluid's linear response to a metric perturbation \cite{Son:kappa} and establishes a
connection between the transport coefficient $\kappa$ and the retarded thermal correlator of the
energy momentum tensor $T_{ij}$ in momentum space, 
\begin{align}
\label{eq:t12t12_ret}
  G^\mathrm{R}(x,y) &=
  \left<\left[T_{12}(x),T_{12}(y)\right]\theta(x_0-y_0)\right>,\\
  G^\mathrm{R}(\omega, \vec{q})& = \int\limits_{-\infty}^\infty \dd t \, \dd x^3 ~ \e^{-\ii (\omega t - q_i x_i)}
G^\mathrm{R}(x,0).
\end{align}
The transport coefficient $\kappa$ is identified as the leading low momentum coefficient at zero
frequency with momentum aligned in $z$-direction, $\vec{q}=(0,0,q_3)$
\cite{Son:kappa,Romatschke:sumrules},
\begin{align}
 G^\mathrm{R} (\omega=0, \vec{q}) &= G(0) + \frac{\kappa}{2}\abs{\vec{q}}^2
				    + \mathcal{O}(\abs{\vec{q}}^4).
 \label{eq:GR}
\end{align}
While the retarded correlator is a real time quantity, 
it is related by analytic continuation to the
Euclidean correlator 
\begin{align}
 \label{eq:t12t12_eucl}
 G^\mathrm{E}(x,y)&=\left<T_{12}(x)T_{12}(y)\right>,\\
  G^\mathrm{E}(\ii\omega_n, \vec{q})&= \int\limits_0^{1/T}\dd \tau \int\limits_{-\infty}^\infty \dd x^3 ~
\e^{-\ii (\omega_n \tau + q_i x_i)} G^\mathrm{E}(x,0),
\end{align}
with the discrete Matsubara frequencies $\omega_n=n2\pi T$, $n\in\mathbb{Z}$. This can be seen by
writing both correlators in their spectral representation
\begin{align}
G^\mathrm{R}(\omega, \vec{q}) &= \ii \int\limits_{-\infty}^\infty \frac{\dd
\omega^\prime}{2\pi} ~ \frac{\rho(\omega^\prime, \vec{q})}{\omega-\omega^\prime+\ii
\eta}, \\
G^\mathrm{E}(\ii \omega_n, \vec{q}) &= \int\limits_{-\infty}^\infty \frac{\dd
\omega}{2\pi} ~ \frac{\rho(\omega, \vec{q})}{\omega - \ii \omega_n}.
\end{align}
Appropriate boundary conditions render the analytic continuation unique \cite{lebellac:tft},
\begin{align}
G^\mathrm{R}(\omega,\vec{q})=G^\mathrm{E}(\omega+i\eta, \vec{q}).
\end{align}
For vanishing frequency $\omega=0$ this can be written \cite{Meyer:review}
\begin{align}
 G^\mathrm{R}(\omega=0, \vec{q}) = G^\mathrm{E}(\omega=0, \vec{q}) + B.
\end{align}
The contact term $B$ arises from the missing commutator in the definition of the Euclidean
correlator \eqref{eq:t12t12_eucl} compared to its retarded analogue \eqref{eq:t12t12_ret} and
corresponds to the correlator evaluated at equal spacetime points, $\sim T_{12}(0)T_{12}(0)$. An
investigation of this contact term $B$ by an operator product expansions shows that it is momentum
independent \cite{Kohno:contact}. Hence equation \eqref{eq:GR} can be rewritten
\begin{align}
 G^\mathrm{E} (\omega=0,\vec{q}) = G^\prime(0) + \frac{\kappa}{2}\abs{\vec{q}}^2
			  + \mathcal{O}(\abs{\vec{q}}^4),
 \label{eq:GE}
\end{align}
where we have absorbed the constant $G(0)$ and the contact term $B$ into a new constant
$G^\prime(0)\equiv G(0)-B$.

The transport coefficient $\kappa$ can now be obtained as the slope of the low momentum correlator
$G^\mathrm{E}(q^2)$, which provides a possibility for a direct 
determination using
lattice QCD. This is in contrast to computations of the shear viscosity \cite{Meyer:eta} or heat
conductivity \cite{Aarts:sigma, Francis:sigma}. These are true dynamical quantities
which cannot be related to Euclidean correlators without non-trivial
analytic continuation. Their determination
by lattice calculations thus requires additional input, e.g. an ansatz for the spectral function or
the maximum entropy method.

So far the discussion was completely general. We now specify to Yang-Mills theory and its energy
momentum tensor
\cite{Meyer:review}
\begin{subequations}
  \begin{align}
    T_{\mu\nu} &= \theta_{\mu\nu} + \frac{1}{4}\delta_{\mu\nu} \theta, \label{eq:TmunuWithTrace} \\
    \theta_{\mu\nu} &= \frac{1}{4}\delta_{\mu\nu}F_{\alpha\beta}^aF_{\alpha\beta}^a -
    F_{\mu\alpha}^a F_{\nu\alpha}^a  \label{eq:Tmunu}
  \end{align}        
\end{subequations}
where $F_{\mu\nu}^a$ corresponds to the field strength tensor. The term
$\theta=\beta(g)/(2g)F_{\alpha\beta}^aF_{\alpha\beta}^a$ includes the renormalisation group function
$\beta(g)$ and corresponds to the trace anomaly, which is caused by breaking of scale invariance.
Since the transport coefficient $\kappa$ is defined in the shear channel,
$\left<T_{12}T_{12}\right>$, it does not enter the computation.

Equation \eqref{eq:GE} has been evaluated perturbatively in pure gluodynamics in the ideal gas
limit, i.e.~at vanishing coupling, with the result
\cite{Romatschke:sumrules, Moore:kappa}
\begin{align}
 \kappa = \left(\colourN^2-1\right)\frac{T^2}{18}.
 \label{eq:kappa_continuous}
\end{align}

\section{\texorpdfstring{Computation of $\kappa$ in lattice QCD}{Computation of kappa in lattice
QCD}}
\label{sec:lattice}
In this section we describe the discretisation of the action and the energy-momentum tensor and
explain the need for renormalisation. Furthermore, we calculate $\kappa$ in lattice perturbation
theory and compare with the continuum result (\ref{eq:kappa_continuous}).

\subsection{Lattice framework}
\label{subsec:setup}
We employ Wilson's Yang-Mills action on an anisotropic lattice with different lattice spacings in
temporal and spatial direction, $a_\sigma$ and $a_\tau$, respectively,
\begin{align}
      S[U]  = \frac{\beta}{\colourN} \RE\Tr \left[ \frac{1}{\xi_0}
	      \sum\limits_{x, i<j} \left(1-U_{ij}(x)\right)
	      + \xi_0 \sum\limits_{x, i} \left(1-U_{i0}(x)\right) \right]
\end{align}
with lattice coupling $\beta={2\colourN}/{g^2}$ and plaquette variables $U_{\mu\nu}$. The bare
anisotropy $\xi_0$ gets renormalised to the actual anisotropy $\xi={a_\sigma}/{a_\tau}$,
\begin{align}
 \eta(\beta,\xi) = \frac{\xi}{\xi_0(\beta,\xi)}. 
\end{align}
We take the numerical evaluation of the renormalisation factor from \cite{Klassen:aniso}. The scale
is set for a specific value of the anisotropy, $\xi=2$, by comparison of the string tension from
the lattice $\sqrt{\sigma_\mathrm{L}}$ \cite{Namekawa:aniso} to its experimental value
$\sqrt{\sigma_\mathrm{exp}}=440\,\mbox{MeV}$ \cite{Allton:stringtension}. The spatial lattice
spacing follows from
\begin{align}
 a_\sigma^{-1}=\frac{\sqrt{\sigma_\mathrm{exp}}}{\sqrt{\sigma_\mathrm{L}}}.
\end{align}

As will be discussed in section \ref{subsec:renorm}, the discretised energy-momentum tensor requires
multiplicative renormalisation due to the reduced translational invariance on the lattice. For this
purpose it is favourable to express the correlator \eqref{eq:t12t12_eucl} in terms of diagonal
elements instead of nondiagonal ones. This is achieved by rotating the lattice by ${\pi}/{4}$ in
the plane of the corresponding channel, i.e.~the $(1,2)$-plane for $\vec{q}=(0,0,q_3)$. As
shown in appendix \ref{app:cubic} the trace anomaly $\theta$ does not enter the transformed
correlator,
although it includes diagonal elementes of the energy-momentum tensor
\cite{Karsch:cubic},
\begin{align}
 \left< T_{12}(x)T_{12}(y)\right> = \frac{1}{2} \Big[\left<
\theta_{11}(x)\theta_{11}(y)\right> - \left<\theta_{11}(x)\theta_{22}(y)\right>\Big].
 \label{eq:cubic}
\end{align}
Additionally, temporal and spatial elements of the energy-momentum tensor require separate
renormalisation factors $Z_\tau$ and $Z_\sigma$ on an anisotropic lattice. The diagonal
energy-momentum tensor elements in the clover discretisation read
\begin{align}
 a_\sigma^3a_\tau \theta_{ii}(x) & = \frac{\beta}{128 N_{\mathrm{c}}} \RE \Tr \left[
Z_\tau(\beta,\xi) \theta_{ii}^\tau(x) + Z_\sigma(\beta,\xi)\theta_{ii}^\sigma \right],
\label{eq:tmunu_clover}
\end{align}
where the bare elements are given by
\begin{subequations}
 \label{eq:tmunu_sep}
 \begin{align}
 \theta_{ii}^\tau(x) &= \xi_0 \widehat{F}_{0i}^2(x)  - \xi_0 \sum_{k\neq i} \widehat{F}_{k0}^2(x),
\\
 \theta_{ii}^\sigma(x) &= - \frac{1}{\xi_0} \sum_{\begin{subarray}{1}k, j\neq i \\ k<j
\end{subarray}}
 \widehat{F}_{kj}^2(x) + \frac{1}{\xi_0} \sum_k \widehat{F}_{ki}^2(x).
 \end{align}
\end{subequations}
The clover plaquette \cite{Luscher:clover} consists of four ordinary plaquettes (see figure
\ref{fig:clover}) and is given by
\begin{align}
  \widehat{F}_{\mu\nu}(x) &= Q_{\mu\nu}(x) - Q_{\nu\mu}(x), \\
  Q_{\mu\nu}(x) &\equiv \frac{1}{4}\left[U_{\mu\nu}(x)+U_{\nu-\mu}(x)
				+U_{-\mu-\nu}(x)+U_{-\nu\mu}(x) \right].
\end{align}
In contrast to an implementation with simple plaquette terms \cite{Karsch:cubic} the clover version
has reduced discretisation errors and  an improved signal-to-noise ratio \cite{Meyer:ntau},
cf.~figure \ref{fig:clover_better}.

\begin{figure}
 \begin{minipage}[b]{.4\textwidth}
  \centering
  \begin{tikzpicture}
   \begin{scope}[scale=3]
     \draw[xstep=.5, ystep=.3, gray] (-.15,-.15) grid (1.65,1.05);
     \draw[-,thick,blue] (0,0) -- (0,-.1);
     \draw[-,thick,blue] (0,-.1) -- (.5,-.1);
     \draw[-,thick,blue] (.5,-.1) -- (.5,0);
     \node[below] (n) at (.25, -.1) {\color{blue}\footnotesize  $a_\sigma = \xi a_\tau$};
     \draw[-,thick,blue] (0,.9) -- (-.1,.9);
     \draw[-,thick,blue] (-.1,.9) -- (-.1,.6);
     \draw[-,thick,blue] (-.1,.6) -- (0,.6);
     \node[left] (n) at (-.1, .75) {\color{blue}\footnotesize  $a_\tau$};
     \filldraw (0,0) circle (1.pt);
     \filldraw (.5,0) circle (1.pt);
     \filldraw (1,0) circle (1.pt);
     \filldraw (1.5,0) circle (1.pt);
     \filldraw (0,.3) circle (1.pt);
     \filldraw (.5,.3) circle (1.pt);
     \filldraw (1,.3) circle (1.pt);
     \filldraw (1.5,.3) circle (1.pt);
     \filldraw (0,.6) circle (1.pt);
     \filldraw (.5,.6) circle (1.pt);
     \filldraw [blue] (1,.6) circle (1.pt);
     \filldraw (1.5,.6) circle (1.pt);
     \filldraw (0,.9) circle (1.pt);
     \filldraw (.5,.9) circle (1.pt);
     \filldraw (1,.9) circle (1.pt);
     \filldraw (1.5,.9) circle (1.pt);
      \draw[thick, blue, rounded corners=1pt] (1.1, .65) --  (1.45,.65) -- (1.45, .85) -- (1.05,.85) -- (1.05, .65) -- (1.4,.65);
      \draw[->,thin, blue] (1.29, .65) -- (1.30, .65);
      \draw[->,thin, blue] (1.45, .74) -- (1.45, .75);
      \draw[->,thin, blue] (1.21, .85) -- (1.20, .85);
      \draw[->,thin, blue] (1.05, .72) -- (1.05, .70);
      \draw[thick, blue, rounded corners=1pt] (1.1, .55) --  (1.45,.55) -- (1.45, .35) -- (1.05,.35) -- (1.05, .55) -- (1.4,.55);
      \draw[->,thin, blue] (1.27, .55) -- (1.26, .55);
      \draw[->,thin, blue] (1.45, .48) -- (1.45, .49);
      \draw[->,thin, blue] (1.29, .35) -- (1.30, .35);
      \draw[->,thin, blue] (1.05, .42) -- (1.05, .41);
      \begin{scope}[shift = {(-.5,0)}]
      \draw[thick, blue, rounded corners=1pt] (1.1, .65) --  (1.45,.65) -- (1.45, .85) -- (1.05,.85) -- (1.05, .65) -- (1.4,.65);
      \draw[->,thin, blue] (1.29, .65) -- (1.30, .65);
      \draw[->,thin, blue] (1.45, .75) -- (1.45, .76);
      \draw[->,thin, blue] (1.21, .85) -- (1.20, .85);
      \draw[->,thin, blue] (1.05, .72) -- (1.05, .70);
      \end{scope}
      \begin{scope}[shift ={(-.5,0)}]
     \draw[thick, blue, rounded corners=1pt] (1.1, .55) --  (1.45,.55) -- (1.45, .35) -- (1.05,.35) -- (1.05, .55) -- (1.4,.55);
      \draw[->,thin, blue] (1.27, .55) -- (1.26, .55);
      \draw[->,thin, blue] (1.45, .48) -- (1.45, .49);
      \draw[->,thin, blue] (1.29, .35) -- (1.30, .35);
      \draw[->,thin, blue] (1.05, .42) -- (1.05, .41);
      \end{scope}
      \node[right] (n) at (1.5, .75) {\color{blue} $Q_{\mu\nu}(x)$};
    \end{scope}
   \end{tikzpicture}
  \captionof{figure}{Illustration of the clover plaquette
	  on an anisotropic lattice.}
  \label{fig:clover}
 \end{minipage}%
 \qquad
 \begin{minipage}[b]{.55\textwidth}
  \centering
  \include{figures/Clover_better}
  \vspace{-0.5cm}
  \captionof{figure}{Computation of $\theta_{11}$
	    on an isotropic $6\times 16^3$ lattice for $\beta=7.1$. We compare
	    the clover and plaquette discretisations.}
  \label{fig:clover_better}
 \end{minipage}
\end{figure}

\subsection{\texorpdfstring{Relation of $\kappa$ to the lattice correlator}{Relation of kappa to
the lattice correlator}}
In order to extract $\kappa$ numerically from equation \eqref{eq:GE}, we compute the Euclidean
correlator of the energy-momentum tensor within the lattice framework and perform a Fourier
transform to momentum space with vanishing frequency. The determination requires the momenta to be
aligned orthogonally to the studied channel of the energy-momentum tensor, i.e.~$\vec{q}=(0,0,q_3)$
for $T_{12}$. This is also the case for the corresponding Kubo formula \cite{Meyer:review}. Thus the
correlator in momentum space is given by
\begin{align}
 a_\sigma^3 a_\tau G^\mathrm{E}(q_3) = \frac{1}{V}\sum\limits_{x,y} \e^{-\ii q_3(x_3-y_3)}
\left<T_{12}(x)T_{12}(y)\right>.
 \label{eq:GE_lat}
\end{align}
Additionally, we include the channels $T_{13}$ and $T_{23}$ with corresponding momenta in our
analysis, since rotational invariance allows to average over all three channels.

We need small momenta compared to temperature, which sets the relevant scale, i.e.~${q_i}/{T}<1$.
With the discretised versions of temperature and momenta
\begin{align}
 T=\frac{1}{a_\tau N_\tau}, \qquad q_i=\frac{2\pi}{a_\sigma N_\sigma}n_i, \qquad n_i=0,1,\dots,N_\sigma-1
\end{align}
we have for the ratio on the lattice
\begin{align}
 \frac{q_i}{T} = \frac{2\pi N_\tau}{\xi N_\sigma}n_i < 1.
 \label{eq:constraint}
\end{align}
The temporal lattice extent $N_\tau$ is fixed by the temperature and lattice spacing. In order to
fit the transport coefficient $\kappa$ to equation \eqref{eq:GE}, we need at least three different
momenta satisfying this constraint \eqref{eq:constraint}. Thus the simulation requires large spatial
lattice extents $N_\sigma$, which makes the calculation costly. This can be partly moderated by
working with anisotropic lattices $\xi>1$.

\subsection{\texorpdfstring{Lattice perturbation theory}{
Lattice perturbation theory}}
\label{subsec:lpt}
In order to estimate lattice artefacts and check our numerics, we first compute the transport
coefficient $\kappa$ in lattice perturbation theory on a lattice with anisotropy $\xi$ in the case
of vanishing coupling $(g=0)$. Definitions of relevant quantities and intermediate results can be
found in appendix \ref{app:lpt}, for an overview see e.g.~\cite{capitani:lpt}.

In the case of vanishing coupling the field strength tensor simplifies to
\begin{align}
 F_{\mu\nu}^a = \partial_\mu^\mathrm{c} A_\nu^a - \partial_\nu^\mathrm{c} A_\mu^a,
 \label{eq:fieldstrength}
\end{align}
where we replace the differential operator by the central difference
\begin{align}
 \partial_\mu^\mathrm{c} A_\nu^a(x) = \frac{1}{a_\mu} \left[A_\nu^a(x+\frac{a_\mu \hat{\mu}}{2})-A_\nu^a(x-\frac{a_\mu
\hat{\mu}}{2}) \right]
\end{align}
and define a lattice spacing $a_\mu$, which is excluded from Einstein's sum convention
\begin{align}
 a_\mu =
 \begin{cases}
  a_\tau \mbox{ for } \mu=0 \\
  a_\sigma \mbox{ for } \mu=1,2,3~.
 \end{cases}
\end{align}
In lattice perturbation theory the dynamical variables are the gauge fields $A_\mu$ and we can plug
the energy-momentum tensor from equation \eqref{eq:Tmunu} together with the field strength tensor
\eqref{eq:fieldstrength} into the correlator \eqref{eq:t12t12_eucl}. Then sixteen terms of the
generalised form
\begin{align}
  C_{i_1 i_2 j_1 j_2 l_1 l_2 m_1 m_2}(x, y) = \left<\partial_{i_1}^\mathrm{c} A_{i_2}^a(x) \partial_{j_1}^\mathrm{c}
A_{j_2}^a(x) \partial_{l_1}^\mathrm{c} A_{l_2}^b(y) \partial_{m_1}^\mathrm{c} A_{m_2}^b(y) \right>
\end{align}
have to be transformed to momentum space. After transforming the individual gauge fields $A_\mu(x)$
to momentum space by \eqref{eq:FT_Amu}, we apply Wick's theorem using the free gauge field
propagator \eqref{eq:latticepropagator}. Because of translational invariance it is sufficient to
consider $y=0$ or
$C_{i_1 i_2 j_1 j_2 l_1 l_2 m_1 m_2}(x, 0)$, and we obtain
\begin{align}
 C_{i_1 i_2 j_1 j_2 l_1 l_2 m_1 m_2}(\omega, \vec{q}) &= (N_{\mathrm{c}}^2-1)
  \sfreqq{k}\frac{\latk_{l_1}(\widetilde{k+q})_{m_1}} {\latk^2 (\widetilde{k+q})^2} \notag \\
  &\quad \times \Big[ \delta_{i_2 l_2}\delta_{j_2 m_2} \latk_{i_1} (\widetilde{k+q})_{j_1} + \delta_{i_2m_2}
  \delta_{j_2 l_2} \latk_{j_1} (\widetilde{k+q})_{i_1} \Big]
\end{align}
with the lattice momenta $\latq$, $\latk$ as defined in appendix \ref{app:lpt}.
Evaluating the correlator \eqref{eq:t12t12_eucl} and aligning the outer momentum to $q=(0,0,0,q_3)$
we find for its Fourier transform
\begin{align}
 G^\mathrm{E}(q) &= \left(\colourN^2-1\right)\sfreqq{k} \frac{1}{\latk^2(\widetilde{q+k})^2}
\Bigg\{4\latk_x^2\latk_y^2 -
          2\latk(\widetilde{q+k})(\latk_x^2+\latk_y^2)+ \latk^2\latk_x^2 \notag \\
      &\qquad\qquad\qquad\qquad\qquad\qquad\,  + (\widetilde{q+k})^2\latk_y^2 +
\left[\latk(\widetilde{q+k})\right]^2\Bigg\}.
\label{eq:integral}
\end{align}
We perform the finite
Matsubara sums by the residue theorem using the formula
\cite{kaste:matsubara}
\begin{align}
  \frac{1}{N_\tau}\sum_{n=1}^{N_\tau} g(z) = -\sum_i \frac{\operatorname{Res}_{\bar{z}_i}\left(
  \frac{1}{z}g(z) \right)}{\bar{z}_i^{N_\tau}-1}
\end{align}
and list the results for the individual terms in appendix \ref{app:matsubara}. As will be described
in section \ref{subsec:renorm} we subtract the temperature independent vacuum part to avoid
ultraviolet divergences. The three-momentum integration can be performed after expanding the
integrals around the continuum limit. This step extends the integration measure to infinite volume
$\left[-\nicefrac{\pi}{a},\nicefrac{\pi}{a}\right]^3 \rightarrow \mathbb{R}^3$ and produces
correction terms in small lattice spacings $a_\sigma$. Together with the expansion in small momenta
$q_3$ the remaining integrals can be solved analytically and one finds for the different terms
\begin{subequations}
\label{eq:lat_integrals}
 \begin{align}
  \sfreqq{k} \frac{4 \latk_x^2\latk_y^2}{\latk^2(\widetilde{q+k})^2} &= \frac{\pi^2}{45(a_\tau N_\tau)^4} +
        \frac{\pi^4a_\sigma^2}{(a_\tau N_\tau)^6}\left(\frac{1}{135}+\frac{5}{189\xi^2}\right)\notag
\\
       &\quad -\frac{q^2}{72(a_\tau N_\tau)^4} + \frac{\pi^2a_\sigma^2q^2}{(a_\tau N_\tau)^4}
            \left(-\frac{1}{1440}-\frac{13}{4320\xi^2}\right), \\
  -\sfreqq{k} \frac{2\latk (\widetilde{q+k}) (\latk_x^2+\latk_y^2)}{\latk^2(\widetilde{q+k})^2}&=
      -\frac{2\pi^2}{45(a_\tau N_\tau)^4} - \frac{\pi^4a_\sigma^2}{(a_\tau
      N_\tau)^6}\left(\frac{2}{189}+\frac{2}{189\xi^2}\right) \notag \\
    &\quad +\frac{q^2}{12(a_\tau N_\tau)^2} - \frac{\pi^2a_\sigma^2q^2}{(a_\tau N_\tau)^4}
        \left(-\frac{1}{2160}-\frac{13}{720\xi^2} \right), \\
  \sfreqq{k} \frac{\latk_x^2}{(\widetilde{q+k})^2} = \sfreqq{k} \frac{\latk_y^2}{\latk^2} &= \frac{\pi^2}{90(a_\tau
N_\tau)^4} + \frac{\pi^4 a_\sigma^2}{(a_\tau N_\tau)^6} \left(\frac{1}{378}+\frac{1}{378\xi^2}\right)
\displaybreak[0], \\
\sfreqq{k} \frac{\left[\latk(\widetilde{q+k})\right]^2}{\latk^2(\widetilde{q+k})^2} &= -\frac{q^2}{24(a_\tau N_\tau)^2}
+ \frac{\pi^2a_\sigma^2 q^2}{(a_\tau N_\tau)^4}\left(-\frac{17}{4320}+\frac{11}{1440\xi^2}\right).
 \end{align}
\end{subequations}
For fixed temperature $T=\left(a_\tau N_\tau\right)^{-1}$ we can rewrite the dependence on lattice
spacings $a_\tau$ and $a_\sigma$ as a dependence on the temporal lattice extent $N_\tau$ and the
anisotropy $\xi=a_\sigma/a_\tau$. Combining the results of \eqref{eq:lat_integrals} we obtain the
following expression for the dimensionless energy-momentum tensor correlator in momentum space
\begin{align}
 \frac{G^E(q)}{T^4} = &\, (\colourN^2-1) \Bigg\{ \frac{\pi^4}{N_\tau^2} \left(\frac{2\xi^2}{945} + \frac{4}{189}\right) \notag\\
		      &\qquad\qquad~\, + \frac{q^2}{T^2} \left[ \frac{1}{36} + \frac{\pi^2}{N_\tau^2} \left( -\frac{\xi^2}{240} + \frac{49}{2160} \right) \right] \Bigg\} + \mathcal{O}\left(q^4,N_\tau^{-4}\right),
\end{align}
from which we identify the dimensionless transport coefficient $\kappa/T^2$ as
\begin{align}
\frac{\kappa}{T^2}= (\colourN^2-1)\left[\frac{1}{18} + \frac{\pi^2}{N_\tau^2}
\left(-\frac{\xi^2}{120} + \frac{49}{1080}\right)\right] +
\mathcal{O}\left(q^4,N_\tau^{-4}\right).
\label{eq:kappalpt}
\end{align}
At fixed temperature the continuum limit $a_\mu\rightarrow 0$ is performed by taking
$N_\tau\rightarrow\infty$, where we reproduce the result of equation
\eqref{eq:kappa_continuous}.

Although the computation has been performed in the ideal gas limit and thus lacks corrections in the
coupling $g$, it may serve as a check of our numerics at high temperatures and 
helps to
estimate the size of cut-off effects. The computed correction in the inverse temporal lattice extent
suggests an anisotropy of $\xi\approx2.33$ in order to eliminate leading order lattice artefacts. In
the case of other values for the anisotropy we can determine the required temporal lattice extent to
decrease the leading discretisation error below a desired treshold. 
As stated in section \ref{sec:results} we use $\xi=2$
in order to use previous results for the scale setting. Thus a temporal lattice extent of $N_\tau
\geq 6$ is required in order to reduce the leading lattice artefacts below $10\%$ in the ideal gas
limit. Note that an anisotropy larger than $\xi>2.33$ causes a quadratic increase of the lattice
artefacts, though it would milden the constraint \eqref{eq:constraint}.

\subsection{Renormalisation}
\label{subsec:renorm}
The correlator defined in \eqref{eq:t12t12_eucl} suffers from ultraviolet divergences. Although they
become finite on the lattice, we have to correct the correlator by additive renormalisation.
Therefore we subtract the vacuum part, which is defined as the correlator computed at vanishing
temperature, from the measured correlator. We define a new vacuum corrected expectation value by
\begin{align}
 \left<\mathcal{O}\right> = \left<\mathcal{O}\right>_T -
  \left<\mathcal{O}\right>_{T_\mathrm{vac}},
 \label{eq:renorm_additive}
\end{align}
where $\left<\mathcal{O}\right>_T$ is an observable evaluated at a given temperature $T$ and
$\left<\mathcal{O}\right>_{T_\mathrm{vac}}$ its vacuum contribution, i.e. evaluated at vanishing
temperature $T_\mathrm{vac}=0$.

The energy-momentum tensor is the Noether current corresponding to translational invariance. In the
continuum it is protected from renormalisation by Ward-identities \cite{Caracciolo:tmunu}. However,
on the lattice translations only form a discrete symmetry group and thus multiplicative
renormalisation becomes necessary. (The lattice perturbation theory computation in section
\ref{subsec:lpt} does not require multiplicative renormalisation because it is the non-interacting
case).

For an isotropic lattice the finite renormalisation factor only depends on the lattice coupling
$\beta$ whereas on an anisotropic lattice it also depends on the anisotropy $\xi$. Additionally,
temporal and spatial direction \eqref{eq:tmunu_sep} require separate renormalisation factors
$Z_\sigma(\beta, \xi)$ and $Z_\tau(\beta, \xi)$. Then the renormalised energy-momentum tensor in the
diagonal channel reads
\begin{align}
 \theta_{ii} & = Z_\tau(\beta, \xi) \left[ \theta_{ii}^\tau +
 \frac{Z_\sigma(\beta, \xi)}{Z_\tau(\beta, \xi)} \theta_{ii}^\sigma\right].
 \label{eq:renorm}
\end{align}
Applying the cubic symmetry \eqref{eq:cubic} we rewrite the correlator \eqref{eq:GE_lat} using the
above notation and find
\begin{align}
 a_\sigma^3 a_\tau G^\mathrm{E}(q_3) &= \frac{1}{2V}\sum\limits_{x,y} \e^{-\ii
		q_3(x_3-y_3)} \big[Z_\tau^2 G_{0}^\tau(x,y)
				+ Z_\tau Z_\sigma G_{0}^{\tau\sigma}(x,y) \notag\\
	&\qquad\qquad\qquad\qquad\quad~+ Z_\sigma^2 G_{0}^\sigma(x,y) \big]
\end{align}
with the newly defined bare correlators
\begin{subequations}
\label{eq:bare_correlators}
 \begin{align}
  G_{0,T}^\tau(x,y) &\equiv \left<\theta_{11}^\tau(x)\theta_{11}^\tau(y)
			   - \theta_{11}^\tau(x)\theta_{22}^\tau(y)\right>_T \\
  G_{0,T}^{\tau\sigma}(x,y) &\equiv \left<\theta_{11}^\tau(x)\theta_{11}^\sigma(y)
			   + \theta_{11}^\sigma(x)\theta_{11}^\tau(y)
			   - \theta_{11}^\tau(x)\theta_{22}^\sigma(y)
			   - \theta_{11}^\sigma(x)\theta_{22}^\tau(y)\right>_T \\
  G_{0,T}^\sigma(x,y) &\equiv \left<\theta_{11}^\sigma(x)\theta_{11}^\sigma(y)
			   - \theta_{11}^\sigma(x)\theta_{22}^\sigma(y)\right>_T,
 \end{align}
\end{subequations}
and their vacuum subtracted versions
\begin{align}
 G_0^i(x,y)= G_{0,T}^i(x,y) - G_{0,T_\mathrm{vac}}^i(x,y),
              \qquad i \in \left\{\tau,\tau\sigma,\sigma\right\}.
\label{eq:renorm_additive_corr}
\end{align}

Performing the renormalisation procedure we need the ratio ${Z_\sigma(\beta, \xi_0)}/{Z_\tau(\beta,
\xi_0)}$ and the absolute scale $Z_\tau(\beta, \xi_0)$. The former can be obtained from
renormalisation group invariant quantities \cite{Meyer:renorm}. To this end one introduces three
differently sized lattices
\begin{align}
  \left< \mathcal{O} \right>_1 ~& \widehat{=}~ 2L\times L\times L\times L, \qquad
  \left< \mathcal{O} \right>_2 ~ \widehat{=}~ L\times 2L\times L\times L, \notag\\
  \left< \mathcal{O} \right>_3 ~& \widehat{=}~ L\times L\times 2L\times L, \qquad
  \left< \mathcal{O} \right>_4 ~ \widehat{=}~ L\times L\times L\times 2L,
 \label{eq:renorm_lattsizes}
\end{align}
and the renormalisation group invariant quantities
\begin{align}
  F_1 =L^4\left<T_{00}\right>_1, \quad F_2 =L^4\left<T_{11}\right>_2, \quad
  F_3 =L^4\left<T_{22}\right>_3, \quad F_4 =L^4\left<T_{33}\right>_4.
 \label{eq:renorm_simulations}
\end{align}
Since the renormalisation factors do not depend on the temperature, all directions are symmetric and
it follows
\begin{align}
 F_1=F_2, \qquad F_1=F_3, \qquad F_1=F_4.
 \label{eq:renorm_equations}
\end{align}
Applying equation \eqref{eq:renorm} one can solve for the ratio of renormalisation factors. For
instance the equation $F_1=F_2$ translates to
\begin{align}
 \frac{Z_\sigma(\beta, \xi)}{Z_\tau(\beta, \xi)} =
\frac{\left<\theta_{00}^\tau\right>_1-\left<\theta_{11}^\tau\right>_2}{\left<\theta_{11}
^\sigma\right>_2-\left<\theta_{00 }^\tau\right>_1},
\end{align}
where the expectation values are computed by lattice simulations of \eqref{eq:renorm_simulations}.
We compute the ratio ${Z_\sigma(\beta, \xi_0)}/{Z_\tau(\beta,\xi_0)}$ from all three equations in
\eqref{eq:renorm_equations} and average the results. The simulations have to be performed for every
lattice coupling $\beta$ and anisotropy $\xi$.

\begin{figure}
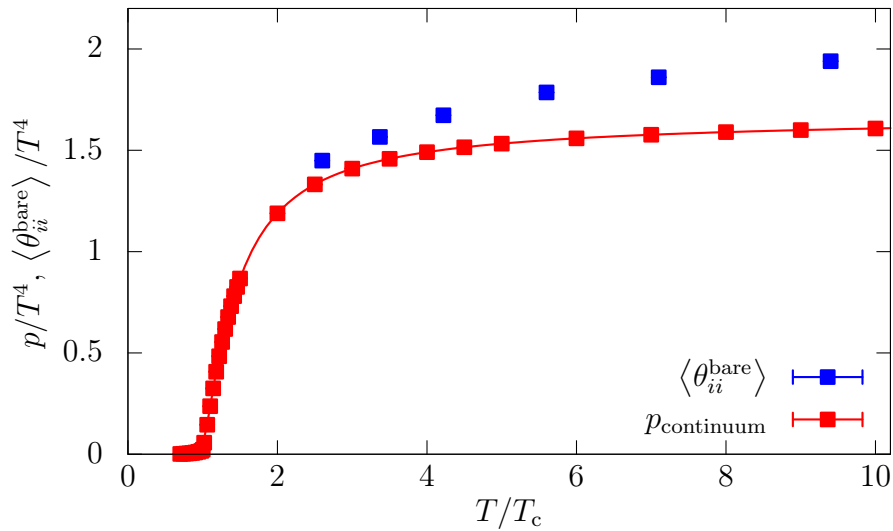

 \centering
 \include{figures/P_T}
 \caption{Comparison of the not multiplicatively renormalised energy-momentum tensor
          $\left<\theta_{ii}^\mathrm{bare}\right>/T^4$ for $N_\tau=6$ and $\xi=2$ to the continuum
          extrapolated pressure ${p}/{T^4}$ from the lattice \cite{Borsanyi:pressure}, where the
          line is obtained by a cubic spline interpolation. The difference between them at a given
          temperature corresponds to the absolute renormalisation factor.}
 \label{fig:renorm}
\end{figure}

We obtain the absolute renormalisation factor by utilising the physical interpretation of the
energy-momentum tensor, whose diagonal spatial elements are equivalent to the pressure
\begin{align}
  \left<\theta_{ii}\right> = p.
\label{eq:Tmunu=p}
\end{align}
The absolute renormalisation factor enters the energy-momentum tensor correlator quadratically.
Therefore the renormalisation procedure is very sensitive to the exact value of the pressure and
encourages us to use a highly precise value for it. For this reason we use the continuum
extrapolated lattice data from \cite{Borsanyi:pressure}. Figure \ref{fig:renorm} illustrates the
difference between the continuum value of the pressure and the not multiplicatively renormalised
energy-momentum tensor. The difference between them at a given temperature corresponds to the
absolute renormalisation factor.

\section{Numerical Results}
\label{sec:results}

\subsection{Numerical setup}
\label{subsec:numerics}
We create the gauge field configurations using the standard heatbath algorithm
\cite{creutz:su2, cabibbo:su3, kennedy:su2} adapted to an anisotropic lattice.
Our implementation is based on the library \texttt{QDP++}
\cite{joo:qdp}.

In order to compute the vacuum part necessary for additive renormalisation, we run extra simulations
with increased temporal lattice extent $N_\tau$. For our fine and 
spatially large lattices
this is very costly. We therefore choose
$T_\mathrm{vac}\approx0.8\,T_\mathrm{c}$, with the critical temperature
$T_\mathrm{c}\approx260\,\mbox{MeV}$ for Yang-Mills theory \cite{Boyd:T_c}. 
For our purposes this
temperature is low enough since firstly the vacuum divergence is temperature 
independent, and secondly it is well known that the pressure or the 
deviation of screening masses from their vacuum values are 
exponentially small in the confined
phase (see
\cite{Boyd:T_c,Borsanyi:pressure,Datta:screening,Maezawa:screening} 
for numerical evidence and \cite{Langelage:screening}
for an analytic explanation).

The set of momenta has to fulfil the constraint \eqref{eq:constraint}, which basically dictates the
simulation parameters. An anisotropy $\xi>1$ benefits this constraint. As discussed in section
\ref{subsec:lpt} a value for the anisotropy of $\xi\approx2.33$ minimizes the first order lattice
corrections. However, we choose an anisotropy of $\xi=2$, which allows to set the scale by
taking the lattice spacing as a function of the lattice coupling $a=a(\beta)$ from
\cite{Namekawa:aniso}. Adjusting the temporal lattice extent to $N_\tau\geq6$ reduces the computed
lattice errors in \eqref{eq:kappalpt} below $10\%$. A numerical analysis of the relevant correlators
in lattice QCD \cite{Meyer:ntau} even suggests values for the temporal lattice extent of
$N_\tau\geq8$.

Extracting the transport coefficient $\kappa$ from \eqref{eq:GE} by performing a linear fit in
$q_3^2$ requires at least three different momenta $q_3$, where the highest momentum still has to
fulfil the constraint \eqref{eq:constraint}. More momenta would be favourable improving the fit's
quality. Thus we choose for the temporal lattice extent $N_\tau=6$ and for the spatial lattice
extent $N_\sigma=120$ at a given anisotropy $\xi=2$. All simulation parameters are listed in table
\ref{tab:params}. In the deconfined phase topological fluctuations 
are suppressed \cite{Lucini:topology} and we expect no difficulties in using very fine
lattices. 
Due to the large computational effort creating gauge fields on $120^3\times
N_\tau$ lattices, we do not exclude any configurations but account for existing correlations by
jackknife error sampling, see e.g. \cite{Muenster:qft}.

The multiplicative renormalisation procedure requires knowledge of the renormalisation factor ratio
${Z_\sigma(\beta, \xi_0)}/{Z_\tau(\beta, \xi_0)}$. As described in section \ref{subsec:renorm} we
determine it from computing the quantities \eqref{eq:renorm_simulations} on lattices
\eqref{eq:renorm_lattsizes} with $L=48$. The simulations must be performed for every lattice
coupling $\beta$ of table \ref{tab:params}. Intermediate results for the computation of the
renormalisation factors are shown in table \ref{tab:tmunu_ratio} and table \ref{tab:tmunu_p} in
appendix \ref{app:results} with reference to run (\romannumeral 1) of table \ref{tab:params}.

\begin{table}
  \centering
      \begin{tabular}{|c|c|c|c|c|}
       \hline
        Run                                             &\romannumeral 1&\romannumeral 2&\romannumeral 3&\romannumeral 4 \\ \hline
       $\beta$                                          & $7.1$         &$7.1$          &$6.68$         &$6.14$ \\
       $N_\tau$                                         & $6$           &$8$            &$6$            &$6$    \\
       $N_\sigma$                                       & $120$         &$120$          &$120$          &$120$  \\
       $N_\tau^\mathrm{vac}$                            & $72$          &$72$           &$42$
  &$24$   \\
       $\xi$                                            & $2$           &$2$            &$2$            &$2$    \\ \hline
       $a_\sigma\,[\mbox{fm}]$                          & $0.026$       &$0.026$        &$0.044$        &$0.094$\\
       $\nicefrac{T}{T_\mathrm{c}}$                     & $9.4$         &$7.1$          &$5.6$          &$2.6$  \\
       $\nicefrac{T_\mathrm{vac}}{T_\mathrm{c}}$        & $0.8$         &$0.8$          &$0.8$
  &$0.7$  \\
       \hline
       \# configurations $T$                            & $500800$              &$434480$               &$403500$               &$542000$       \\
       \# configurations $T_\mathrm{vac}$               & $455000$              &$455000$
  &$429000$               &$421250$       \\
       \hline
     \end{tabular}
 \caption{Simulation parameters for four evaluations of $\kappa$. The lower temperature
$T_\mathrm{vac}$ is required for renormalisation.
}
 \label{tab:params}
\end{table}

\subsection{Comparison to lattice perturbation theory}
\label{sec:results:lpt}
Our first simulation aims at making contact to lattice peturbation theory, section \ref{subsec:lpt}.
The weak coupling regime is reached by increasing the temperature. Adopting the parameters from the
previous section \ref{subsec:numerics} we choose for the lattice coupling $\beta=7.1$, corresponding
to a temperature of $T=9.4\,T_\mathrm{c}$, and a spatial lattice spacing of
$a_\sigma=0.026\,\mbox{fm}$ (see column ({\romannumeral 1}) in table \ref{tab:params}).

Figure \ref{fig:kappa} shows the correlator ${G^\mathrm{E}(q)}/{T^4}$ for five momenta compared to
the result from lattice perturbation theory and table \ref{tab:results_corr} the corresponding data
points. The large errors of the correlator are almost entirely due to the additive renormalisation
procedure. Table \ref{tab:bare_correlators} lists the data of the bare correlators
\eqref{eq:bare_correlators} regarding this simulation, whereas table
\ref{tab:subtracted_correlators} lists the data of the additively renormalised correlators
\eqref{eq:renorm_additive_corr}. The vacuum subtraction causes a significant loss of accuracy.
Computing the pressure by means of the interaction measure \cite{Boyd:T_c} suffers from the same
phenomenon. Thus, we create a large amount of statistics (see table \ref{tab:params}) to provide a
significant signal for the correlators. In terms of error reduction it is highly favourable to
perform the additive renormalisation before the multiplicative one. Otherwise, the propagated errors
entering from the multiplicative renormalisation add to the described loss of precision.

Fitting the datapoints of the correlator to a line
\begin{align}
  \frac{G^\mathrm{E}\left(\frac{q^2}{T^2}\right)}{T^4} = \frac{G^\prime(0)}{T^4} +
                                                \frac{\kappa}{T^2}\frac{q^2}{2T^2}
 \label{eq:linfit}
\end{align}
yields for the y-intercept $G^\prime(0)/T^4=0.69(4)$ and for the transport coefficient
 $\kappa/T^2=0.40(26)$, which is consistent with the leading order lattice perturbation theory
result $\kappa_\mathrm{LPT}/T^2=0.47$. Note that full agreement is not yet expected since at
$T=9.4\,T_\mathrm{c}$ there are still significant corrections due to interactions, i.e.~we are still
far from the ideal gas limit.

\begin{figure}
 \begin{minipage}[b]{0.65\textwidth}
  \centering
  \include{figures/G_k_kappa}
  \vspace{-.5cm}
  \captionof{figure}{Correlator ${G^\mathrm{E}(q)}/{T^4}$ for
	momenta ${q^2}/{T^2}<1$ compared to results from lattice perturbation theory
	(LPT). The slope of the linear fit gives $\kappa/2$.}
  \label{fig:kappa}
 \end{minipage}
 \quad
 \begin{minipage}[b]{0.3\textwidth}
    \centering
     \begin{tabular}{|c|c|}\hline
      $q^2/T^2$	&$G(q)/T^4$	 	\\ \hline
      0.02		&$0.68(6)$	\\
      0.10		&$0.72(6)$	\\
      0.22		&$0.77(6)$	\\
      0.39		&$0.75(6)$	\\
      0.62		&$0.82(6)$	\\\hline
    \end{tabular}
  \vspace{.3cm}
    \captionof{table}{Intermediate numerical results for run ({\romannumeral 1}) of table
  \ref{tab:params}.}
  \label{tab:results_corr}
 \end{minipage}
\end{figure}

\subsection{Temperature dependence}

In principle, the temperature can be varied at fixed $\beta$ and lattice spacing by changing
$N_\tau$, where lower temperature implies larger $N_\tau$. However, due to the constraint on the
momenta from equation \eqref{eq:constraint} this would require a similar increase of the spatial
volume and thus a drastical growth of the numerical effort. Hence the fixed scale approach is not
practical for temperatures approaching the phase transition.
 
We therefore investigate the temperature dependence of $\kappa$ at fixed $N_\tau/N_\sigma$ by
repeating the simulations at various lattice couplings $\beta$. In this case the different
temperatures are evaluated at different lattice spacings, and consequently also different spatial
volumes in physical units. However, since our lattice spacings are all $a_\sigma<0.1\,\mbox{fm}$, we
expect  the lattice artefacts on the temperature dependence of the transport coefficient
$\kappa/T^2$ to be negligible. As a consistency check for this, we also perform simulations at
different temperatures but the same lattice spacings (simulations ({\romannumeral
1}) and ({\romannumeral 2}) in table \ref{tab:params}). 

\begin{figure}[t]
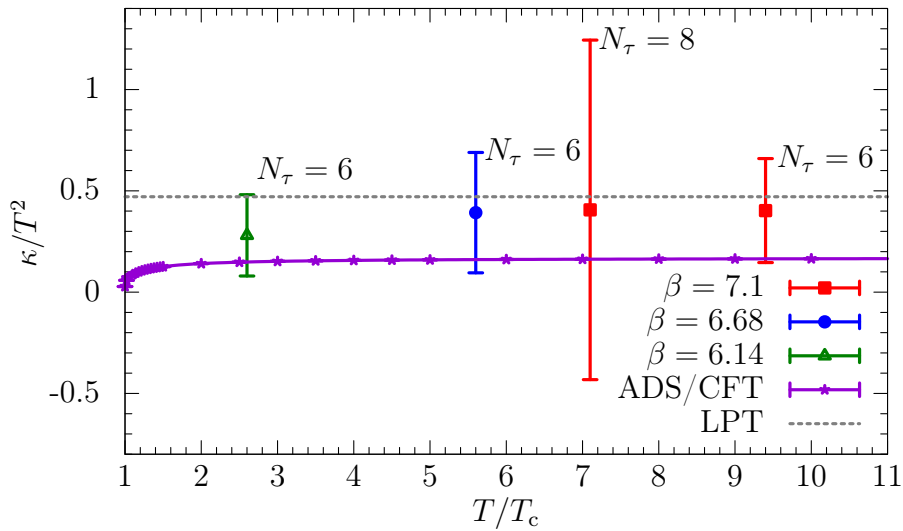

 \centering
 \include{figures/T_kappa}
 \caption{Temperature dependence of the transport coefficient ${\kappa}/{T^2}$. The lines mark
          the result from ADS/CFT correspondence \cite{Son:kappa} and lattice perturbation theory
          \eqref{eq:kappalpt}, respectively.}
 \label{fig:tkappa}
\end{figure}

The results are shown in figure \ref{fig:tkappa}. The datapoint at $T=7.1\,T_\mathrm{c}$ suffers
from large errorbars since the spatial lattice extents have been kept fixed while increasing the
temporal lattice extent $N_\tau$. This corresponds to less momenta fulfilling the constraint
\eqref{eq:constraint} and generates a loss of accuracy in the fit \eqref{eq:linfit}. Within the
errorbars, the values of ${\kappa}/{T^2}$ at $T=9.4\,T_\mathrm{c}$ and $T=7.1\,T_\mathrm{c}$ agree
(c.f.~table \ref{tab:results}) and thus justify the comparison at different lattice spacings and
temperatures.

The numerical values for the transport coefficient $\kappa$ are also summarised in table
\ref{tab:results}. Within errorbars, the temperature dependence of the transport coefficient is
consistent with that of the ideal gas, $\kappa\sim T^2$, which is also the 
prediction of AdS/CFT
\cite{Son:kappa} for the opposite strong coupling limit. 
Assuming this functional dependence, we may increase the accuracy by averaging the data points with
$N_\tau=6$ to give our final result,
\begin{align}
 \kappa^\mathrm{avr} = 0.36(15)T^2.
\end{align}
The prediction from AdS/CFT correspondence for this coefficient is \cite{Romatschke:hydro} 
\begin{align}
 \frac{\kappa}{T^2}=\frac{\eta}{s}\times\frac{s}{\pi T^3},       
\label{adsk}
\end{align}
where $\eta$ is the shear viscosity and $s$ the entropy density. The latter
is proportional to the number of degrees of freedom of the theory, which 
is higher in the SUSY Yang-Mills used for the 
correspondence\footnote{We missed this point in the first version of the 
manuscript and thank the referee and editor for their suggestions.}.
In order to compare with the QCD calculations, we thus use the AdS/CFT results
$\eta/s= 1/4\pi$ and eq.~(\ref{adsk}), but
take the QCD entropy density from a lattice
calculation \cite{Borsanyi:pressure}. The result is about a third of the
perturbative prediction and also consistent with the simulation results. 

\begin{table}[t]
 \centering
     \begin{tabular}{|c|c|c|c|c|}
       \hline
       $\nicefrac{T}{T_\mathrm{c}}$	& $9.4$		&$7.1$		&$5.6$		&$2.6$	\\
       $a_\sigma\,[\mbox{fm}]$		& $0.026$	&$0.026$	&$0.044$	&$0.094$\\
\hline
       $\nicefrac{\kappa}{T^2}$		& $0.40(26)$	&$0.41(84)$	&$0.39(30)$	&$0.28(20)$\\       
       \hline
     \end{tabular}
 \caption{Lattice results for the transport coefficient ${\kappa}/{T^2}$ at different spatial lattice spacings $a_\sigma$ and temperatures ${T}/{T_\mathrm{c}}$.}
 \label{tab:results}
\end{table}

\section{Conclusions}
\label{sec:conclusion}

We have calculated the second order hydrodynamic transport coefficient $\kappa$ 
for the Yang-Mills plasma using lattice perturbation theory 
and Monte Carlo simulations. This is possible because
the retarded correlator of the energy momentum tensor at zero frequency
has a trivial analytic continuation to a corresponding Euclidean correlator.
The transport coefficient parametrises the low momentum behaviour of this
correlator, whose realisation requires large spatial lattice 
directions, making a numerical 
calculation very challenging and thus leaving large
statistical errors. Their main source are the vacuum subtractions
leading to similar problems in calculations of the equation of 
state at low temperatures. One might hope that alternative methods 
avoiding this step \cite{Giusti:poincare} may improve this situation.

In the investigated temperature range $2T_\mathrm{c}<T<10T_\mathrm{c}$ our data
are consistent with
$\kappa\propto T^2$, as predicted both by weak and strong coupling methods.
Because of still large errorbars,
our result also quantitatively covers both the leading order perturbative
as well as the AdS/CFT prediction rescaled by the QCD entropy. This would
suggest that, besides improved simulation methods, next-to-leading order
analytic calculations should be able to give a result with improved
accuracy.

\acknowledgments
We thank H.~Meyer and D.~Rischke for their comments on the manuscript.
This project is supported by the Helmholtz International Center for FAIR 
within the LOEWE program of the State of Hesse. C.S. acknowledges travel support
by the Helmholtz Research School H-QM.
The calculations have been performed on LOEWE-CSC at Goethe-University 
Frankfurt,
we thank the team of administrators for support.

\appendix

\section{Cubic symmetry of the energy momentum tensor}
\label{app:cubic}
The correlator $\left<T_{12}(x)T_{12}(y)\right>$ can be expressed in terms
of diagonal energy-momentum tensor elements by exploiting rotation invariance
\begin{equation}
\langle T'_{12}(x)T'_{12}(y)\rangle=\langle T_{12}(x)T_{12}(y)\rangle\;, 
\end{equation}
on a spatially isotropic lattice (and medium) under rotations by $\alpha=\pi/4$ about the
$z$-direction.
The transformation of a second rank tensor reads
\begin{align}
  T_{\mu\nu}^\prime(x) =  \left(M_z^{-1}\right)_{\mu\alpha}\left(M_z^{-1}\right)_{\nu\beta}
  T_{\alpha\beta}(x),
\end{align}
and the corresponding transformation matrix is given by
\begin{align}
 M_z^{-1}=
 \begin{pmatrix}
  1     &0              &0              &0      \\
  0     &\cos\alpha     &\sin\alpha     &0      \\
  0     &-\sin\alpha    &\cos\alpha     &0      \\
  0     &0              &0              &1
 \end{pmatrix}
= \frac{1}{\sqrt{2}}
\begin{pmatrix}
  \sqrt{2}      &0      &0      &0      \\
  0             &1      &1      &0      \\
  0             &-1     &1      &0      \\
  0             &0      &0      &\sqrt{2}
 \end{pmatrix}.
\end{align}
For the energy-mometum tensor components of interest, this means
\begin{align}
 T_{12}^\prime(x)= \frac{1}{2}\left[T_{22}-T_{11}\right],
\end{align}
where we used $T_{12}=T_{21}$. With the definition of the
energy-momentum tensor \eqref{eq:TmunuWithTrace} we find for the correlator
\begin{align}
 T_{12}^\prime(x)T_{12}^\prime(y) &=
    \frac{1}{4}\left[T_{22}(x)T_{22}(y)-T_{22}(x)T_{11}(y)
        -T_{11}(x)T_{22}(y)+T_{11}(x)T_{11}(y)\right] \notag\\
    & = \frac{1}{4}\left[\theta_{22}(x)\theta_{22}(y) + \theta_{11}(x)\theta_{11}(y)
            -\theta_{22}(x)\theta_{11}(y)-\theta_{11}(x)\theta_{22}(y)\right].
\end{align}
Note that the trace anomaly $\theta$ cancels completely. From rotational invariance follows
\begin{align}
  \left<\theta_{22}(x)\theta_{11}(y)\right> =\left<\theta_{11}(x)\theta_{22}(y)\right>, \qquad
  \left<\theta_{11}(x)\theta_{11}(y)\right> =\left<\theta_{22}(x)\theta_{22}(y)\right>, 
\end{align}
and the correlator expressed in diagonal elements reads
\begin{align}
 \left<T_{12}(x)T_{12}(y)\right> &= \frac{1}{2}\left[
 \left<\theta_{11}(x)\theta_{11}(y)\right> - \left<\theta_{11}(x)\theta_{22}(y)\right>\right].
\end{align}

\section{Definitions in lattice perturbation theory}
\label{app:lpt}
The Fourier transforms of the gauge field $A_\mu$ to momentum space and back are defined by
\begin{subequations}
 \begin{align}
  A_\mu(q) &= a_\sigma^3 a_\tau \sum_{n=1}^{N_\tau}\sum_{\vec{x}}
  \e^{-\ii\left(x+\frac{a_\mu\widehat{\mu}}{2}\right)q}A_\mu(x),\\
  A_\mu(x) &= \sfreqq{q} ~ \e^{\ii\left(x+\frac{a_\mu\widehat{\mu}}{2}\right)q} A_\mu(q) \label{eq:FT_Amu},
 \end{align}
\end{subequations}
where we introduce
\begin{align}
\sfreqq{q} \equiv \frac{1}{a_\tau N_\tau} \sum_{n=1}^{N_\tau}
~ \int\limits_{-\frac{\pi}{a_\sigma}}^{\frac{\pi}{a_\sigma}}\frac{\dd^3q}{(2\pi)^3}.
\end{align}
The shift to the center of the link variables $x+{a_\mu\widehat{\mu}}/{2}$ in the Fourier transform simplifies the
computation. The free gauge field propagator is given by
\begin{align}
  \Delta_{\mu\nu}^{AB}(q) = \frac{1}{\latq^2} \left( \delta_{\mu\nu} - \left( 1-\xi \right) \frac{\latq_\mu
\latq_\nu}{\latq^2}\right)\delta^{AB},
 \label{eq:latticepropagator}
\end{align}
where we use Feynman-'t\,Hooft gauge with $\xi=1$. The momenta in lattice perturbation theory are given by
\begin{align}
 \latq_\mu &= \frac{2}{a_\mu}\sin\left( \frac{a_\mu q_\mu}{2} \right), \\
 \left(\widetilde{k_\mu + q_\mu}\right) &= \frac{2}{a_\mu}\sin\left( \frac{a_\mu (k_\mu + q_\mu)}{2} \right), \\
 \sum_\mu\latq_\mu^{\,2} &= \frac{4}{a_\mu^2}\sum_\mu \sin^2\left( \frac{a_\mu q_\mu}{2} \right)
\end{align}
with $a_0\equiv a_\tau$ and $a_i\equiv a_\sigma$. We do not imply a sum over the index $\mu$.

\section{Results for finite Matsubara sums}
\label{app:matsubara}
The evaluation of the finite Matsubara sums gives with the definitions
\begin{subequations}
 \begin{align}
  E(k_i) &\equiv \xi^{-1}\abs{k_i} - \frac{a_\sigma^2}{24\abs{k_i}}\left[\xi^{-1} \sum_i k_i^4 + \xi^{-3} \abs{k_i}^4
  \right] + \mathcal{O}(a_\sigma^4) \\
  E_1 &\equiv E(k_i) \\
  E_2 &\equiv E(k_i+q_i)
 \end{align}
\end{subequations}
and
\begin{subequations}
 \begin{align}
A &\equiv \frac{1}{\cosh(a_\sigma E_1) - \cosh(a_\sigma E_2)} \\
B &\equiv \frac{1}{\tanh(a_\sigma E_2)} ~ \frac{1}{\e^{a_\sigma N_\tau E_2}-1}
            - \frac{1}{\tanh(a_\sigma E_1)} ~ \frac{1}{\e^{a_\sigma N_\tau E_1}-1} \\
C_1 &\equiv \frac{1}{\sinh(a_\sigma E_1)} \\
C_2 &\equiv \frac{1}{\sinh(a_\sigma E_2)}
 \end{align}
\end{subequations}
the following results
\begin{subequations}
 \begin{align}
  \frac{1}{N_\tau}\sum_{n=1}^{N_\tau} \frac{1}{(\widetilde{k+q})^2} &=
      \frac{a_\tau^2}{2} \left[1+\frac{2}{\e^{a_\sigma N_\tau E_2}-1}\right]C_2 \\
  \frac{1}{N_\tau}\sum_{n=1}^{N_\tau} \frac{1}{\latk^2(\widetilde{k+q})^2} &=
         \frac{a_\tau^4}{4} \Big[C_2 \e^{a_\sigma E_2} - C_1 \e^{a_\sigma E_1}\Big]A
        + \frac{a_\tau^4}{2}AB \\
  \frac{1}{N_\tau}\sum_{n=1}^{N_\tau} \frac{\latk (\widetilde{k+q})}{\latk^2(\widetilde{k+q})^2} &=
        \frac{a_\tau^2}{\e^{a_\sigma N_\tau E_2}-1}C_2 + \frac{a_\tau^4}{2}\left[\latk_i(\widetilde{k_i+q_i})-
\latk_i^2\right]AB \\
\displaybreak[0] \\
  \frac{1}{N_\tau}\sum_{n=1}^{N_\tau} \frac{\left[\latk(\widetilde{k+q})\right]^2}{\latk^2(\widetilde{k+q})^2} &=
      - \frac{a_\tau^2}{\e^{a_\sigma N_\tau E_2}-1}\left[(\widetilde{k_i+q_i}) - \latk_i\right]^2 C_2\notag \\
      & \quad+ \frac{a_\tau^4}{2} \left[\latk_i(\widetilde{k_i+q_i}) - \latk_i^2\right]^2AB  \\
\frac{1}{N_\tau}\sum_{n=1}^{N_\tau} \frac{1}{\latk^2} &=
    \frac{a_\tau^2}{2} \left[1+\frac{2}{\e^{a_\sigma N_\tau E_1}-1}\right]C_1.
 \end{align}
\end{subequations}

\section{Numerical intermediate results}
\label{app:results}
In this section we present numerical intermediate results for run ({\romannumeral 1}) of table
\ref{tab:params}.

\begin{table}[ht]
\centering
\begin{tabular}{|c|c|c|c|c|c|c|}
\hline
$n$	&$G^\tau_{0,T}(q)$	&$G^\tau_{0,T_\mathrm{vac}}(q)$ &$G^\sigma_{0,T}(q)$
&$G^\sigma_{0,T_\mathrm{vac}}(q)$	&$G^{\tau\sigma}_{0,T}(q)$
&$G^{\tau\sigma}_{0,T_\mathrm{vac}}(q)$ \\
\hline
$0$	&$-0.3(6)$	&$2.(2)$		&$0.1(1)$	&$-0.5(4)$	
&$0.03(39)$		&$0.4(1.3)$	\\
$1$	&$0.1982(2)$	&$0.2009(2)$		&$0.04092(4)$	&$0.03768(4)$	
&$0.0283(1)$		&$0.0320(1)$	\\
$2$	&$0.1985(2)$	&$0.2005(2)$		&$0.04092(4)$	&$0.03766(4)$	
&$0.0280(1)$		&$0.0319(1)$	\\
$3$	&$0.1984(2)$	&$0.2003(2)$		&$0.04076(4)$	&$0.03759(4)$	
&$0.0278(1)$		&$0.0314(1)$	\\
$4$	&$0.1983(2)$	&$0.2002(2)$		&$0.04071(4)$	&$0.03743(4)$	
&$0.0273(1)$		&$0.0312(1)$	\\
$5$	&$0.1981(2)$	&$0.2000(2)$		&$0.04059(4)$	&$0.03730(4)$	
&$0.0271(1)$		&$0.0308(1)$	\\
\hline
\end{tabular}  
\caption{Simulation results for the bare correlators $G^\tau_{0,T}$, $G^\sigma_{0,T}$ and
$G^{\tau\sigma}_{0,T}$
and their vacuum parts $G^\tau_{0,T_\mathrm{vac}}$, $G^\sigma_{0,T_\mathrm{vac}}$,
$G^{\tau\sigma}_{0,T_\mathrm{vac}}$ in momentum space for six momentum modes $n$ fulfilling the
constraint \eqref{eq:constraint}.}
\label{tab:bare_correlators}
\end{table}

\begin{table}[ht]
\centering
\begin{tabular}{|c|c|c|c|}
\hline
$n$     &$G^\tau_0(q)$  &$G^\sigma_0(q)$ &$G^{\tau\sigma}_0(q)$ \\
\hline
$0$     &$-3.(3)$       &$0.7(5)$       &$-0.4(1.4)$     \\
$1$     &$-0.0027(3)$   &$0.00324(5)$   &$-0.0037(2)$    \\
$2$     &$-0.0019(3)$   &$0.00327(5)$   &$-0.0040(2)$    \\
$3$     &$-0.0020(3)$   &$0.00317(5)$   &$-0.0035(2)$    \\
$4$     &$-0.0019(3)$   &$0.00329(5)$   &$-0.0039(2)$    \\
$5$     &$-0.0019(3)$   &$0.00329(5)$   &$-0.0036(2)$    \\
\hline
\end{tabular}
\caption{According to \eqref{eq:renorm_additive_corr} vacuum subtracted correlators of table
\ref{tab:bare_correlators}.}
\label{tab:subtracted_correlators}
\end{table}

\begin{table}[ht]
\centering
\begin{tabular}{|c|c|c|c|c|c|c|}
\hline
$i$		&$<\theta_{00}^i>_1$	&$<\theta_{11}^i>_2$ &$<\theta_{22}^i>_3$
&$<\theta_{33}^i>_4$ \\ \hline
$\tau$		&$-1.447172(1)$		&$0.4823904(9)$		&$0.4823868(9)$	
&$0.4823887(8)$		\\
$\sigma$	&$0.6218880(7)$		&$-0.2072959(4)$	&$-0.2072969(5)$
&$-0.2072961(1)$	\\
\hline
\end{tabular}  
\caption{Diagonal energy-momentum tensor elements evaluated on lattices
\eqref{eq:renorm_lattsizes} in order to compute the renormalisation ratio
$Z_\sigma(\beta, \xi)/Z_\tau(\beta, \xi)$.}
\label{tab:tmunu_ratio}
\end{table}

\begin{table}[ht]
\centering
\begin{tabular}{|c|c|c|}
\hline
		&$T$			&$T_\mathrm{vac}$	\\ \hline
$\left<\theta_{11}^\tau\right>$	&$-0.4768093(3)$	&$-0.48239182(8)$	\\
$\left<\theta_{22}^\tau\right>$	&$-0.4768096(3)$	&$-0.48239191(8)$	\\
$\left<\theta_{33}^\tau\right>$	&$-0.4768099(3)$	&$-0.48239181(8)$	\\
\hline
\end{tabular}
\hfill
\begin{tabular}{|c|c|c|}
\hline
			&$T$			&$T_\mathrm{vac}$	\\
\hline
$\left<\theta_{11}^\sigma\right>$		&$0.2100439(1)$		&$0.20729721(4)$	\\
$\left<\theta_{22}^\sigma\right>$		&$0.2100440(1)$		&$0.20729719(4)$	\\
$\left<\theta_{33}^\sigma\right>$		&$0.2100438(1)$		&$0.20729725(4)$	\\
\hline
\end{tabular} 
\caption{Energy-momentum tensor elements required to compute the absolute
renormalisation factor $Z_\tau(\beta, \xi)$ from equivalence to the pressure.}
\label{tab:tmunu_p}
\end{table}


\newpage
\bibliographystyle{JHEP}
\bibliography{mybib}


%
%


\end{document}

%% file: figures/P_T.tex
\begin{tikzpicture}[gnuplot]
\gpsolidlines
\path (0.000,0.000) rectangle (12.090,7.264);
\gpcolor{color=gp lt color border}
\gpsetlinetype{gp lt border}
\gpsetlinewidth{1.00}
\draw[gp path] (1.504,0.985)--(1.684,0.985);
\draw[gp path] (11.537,0.985)--(11.357,0.985);
\node[gp node right] at (1.320,0.985) { 0};
\draw[gp path] (1.504,2.328)--(1.684,2.328);
\draw[gp path] (11.537,2.328)--(11.357,2.328);
\node[gp node right] at (1.320,2.328) { 0.5};
\draw[gp path] (1.504,3.671)--(1.684,3.671);
\draw[gp path] (11.537,3.671)--(11.357,3.671);
\node[gp node right] at (1.320,3.671) { 1};
\draw[gp path] (1.504,5.015)--(1.684,5.015);
\draw[gp path] (11.537,5.015)--(11.357,5.015);
\node[gp node right] at (1.320,5.015) { 1.5};
\draw[gp path] (1.504,6.358)--(1.684,6.358);
\draw[gp path] (11.537,6.358)--(11.357,6.358);
\node[gp node right] at (1.320,6.358) { 2};
\draw[gp path] (1.504,0.985)--(1.504,1.165);
\draw[gp path] (1.504,6.895)--(1.504,6.715);
\node[gp node center] at (1.504,0.677) { 0};
\draw[gp path] (3.471,0.985)--(3.471,1.165);
\draw[gp path] (3.471,6.895)--(3.471,6.715);
\node[gp node center] at (3.471,0.677) { 2};
\draw[gp path] (5.439,0.985)--(5.439,1.165);
\draw[gp path] (5.439,6.895)--(5.439,6.715);
\node[gp node center] at (5.439,0.677) { 4};
\draw[gp path] (7.406,0.985)--(7.406,1.165);
\draw[gp path] (7.406,6.895)--(7.406,6.715);
\node[gp node center] at (7.406,0.677) { 6};
\draw[gp path] (9.373,0.985)--(9.373,1.165);
\draw[gp path] (9.373,6.895)--(9.373,6.715);
\node[gp node center] at (9.373,0.677) { 8};
\draw[gp path] (11.340,0.985)--(11.340,1.165);
\draw[gp path] (11.340,6.895)--(11.340,6.715);
\node[gp node center] at (11.340,0.677) { 10};
\draw[gp path] (1.504,6.895)--(1.504,0.985)--(11.537,0.985)--(11.537,6.895)--cycle;
\node[gp node center,rotate=-270] at (0.246,3.940) {$p/T^4$, $\left<\theta_{ii}^\mathrm{bare}\right>/T^4$};
\node[gp node center] at (6.520,0.215) {$T/T_\mathrm{c}$};
\node[gp node right] at (10.069,2.008) {$\left<\theta_{ii}^\mathrm{bare}\right>$};
\gpcolor{rgb color={0.000,0.000,1.000}}
\gpsetlinetype{gp lt plot 0}
\gpsetlinewidth{2.00}
\draw[gp path] (10.253,2.008)--(11.169,2.008);
\draw[gp path] (10.253,2.098)--(10.253,1.918);
\draw[gp path] (11.169,2.098)--(11.169,1.918);
\draw[gp path] (10.660,6.196)--(10.840,6.196);
\draw[gp path] (10.660,6.196)--(10.840,6.196);
\draw[gp path] (8.488,5.982)--(8.488,5.983);
\draw[gp path] (8.398,5.982)--(8.578,5.982);
\draw[gp path] (8.398,5.983)--(8.578,5.983);
\draw[gp path] (6.922,5.782)--(7.102,5.782);
\draw[gp path] (6.922,5.782)--(7.102,5.782);
\draw[gp path] (5.655,5.478)--(5.655,5.480);
\draw[gp path] (5.565,5.478)--(5.745,5.478);
\draw[gp path] (5.565,5.480)--(5.745,5.480);
\draw[gp path] (4.819,5.190)--(4.819,5.194);
\draw[gp path] (4.729,5.190)--(4.909,5.190);
\draw[gp path] (4.729,5.194)--(4.909,5.194);
\draw[gp path] (4.061,4.877)--(4.061,4.879);
\draw[gp path] (3.971,4.877)--(4.151,4.877);
\draw[gp path] (3.971,4.879)--(4.151,4.879);
\gpsetpointsize{6.40}
\gppoint{gp mark 5}{(10.750,6.196)}
\gppoint{gp mark 5}{(8.488,5.983)}
\gppoint{gp mark 5}{(7.012,5.782)}
\gppoint{gp mark 5}{(5.655,5.479)}
\gppoint{gp mark 5}{(4.819,5.192)}
\gppoint{gp mark 5}{(4.061,4.878)}
\gppoint{gp mark 5}{(10.711,2.008)}
\gpcolor{color=gp lt color border}
\node[gp node right] at (10.069,1.446) {$p_\mathrm{ continuum}$};
\gpcolor{rgb color={1.000,0.000,0.000}}
\draw[gp path] (10.253,1.446)--(11.169,1.446);
\draw[gp path] (10.253,1.536)--(10.253,1.356);
\draw[gp path] (11.169,1.536)--(11.169,1.356);
\draw[gp path] (2.103,0.989)--(2.283,0.989);
\draw[gp path] (2.103,0.989)--(2.283,0.989);
\draw[gp path] (2.142,0.991)--(2.322,0.991);
\draw[gp path] (2.142,0.991)--(2.322,0.991);
\draw[gp path] (2.181,0.994)--(2.361,0.994);
\draw[gp path] (2.181,0.994)--(2.361,0.994);
\draw[gp path] (2.311,0.997)--(2.311,0.998);
\draw[gp path] (2.221,0.997)--(2.401,0.997);
\draw[gp path] (2.221,0.998)--(2.401,0.998);
\draw[gp path] (2.350,1.001)--(2.350,1.003);
\draw[gp path] (2.260,1.001)--(2.440,1.001);
\draw[gp path] (2.260,1.003)--(2.440,1.003);
\draw[gp path] (2.389,1.008)--(2.389,1.009);
\draw[gp path] (2.299,1.008)--(2.479,1.008);
\draw[gp path] (2.299,1.009)--(2.479,1.009);
\draw[gp path] (2.429,1.016)--(2.429,1.018);
\draw[gp path] (2.339,1.016)--(2.519,1.016);
\draw[gp path] (2.339,1.018)--(2.519,1.018);
\draw[gp path] (2.468,1.028)--(2.468,1.030);
\draw[gp path] (2.378,1.028)--(2.558,1.028);
\draw[gp path] (2.378,1.030)--(2.558,1.030);
\draw[gp path] (2.488,1.044)--(2.488,1.046);
\draw[gp path] (2.398,1.044)--(2.578,1.044);
\draw[gp path] (2.398,1.046)--(2.578,1.046);
\draw[gp path] (2.507,1.136)--(2.507,1.141);
\draw[gp path] (2.417,1.136)--(2.597,1.136);
\draw[gp path] (2.417,1.141)--(2.597,1.141);
\draw[gp path] (2.547,1.373)--(2.547,1.379);
\draw[gp path] (2.457,1.373)--(2.637,1.373);
\draw[gp path] (2.457,1.379)--(2.637,1.379);
\draw[gp path] (2.586,1.619)--(2.586,1.624);
\draw[gp path] (2.496,1.619)--(2.676,1.619);
\draw[gp path] (2.496,1.624)--(2.676,1.624);
\draw[gp path] (2.625,1.855)--(2.625,1.861);
\draw[gp path] (2.535,1.855)--(2.715,1.855);
\draw[gp path] (2.535,1.861)--(2.715,1.861);
\draw[gp path] (2.665,2.077)--(2.665,2.082);
\draw[gp path] (2.575,2.077)--(2.755,2.077);
\draw[gp path] (2.575,2.082)--(2.755,2.082);
\draw[gp path] (2.704,2.282)--(2.704,2.287);
\draw[gp path] (2.614,2.282)--(2.794,2.282);
\draw[gp path] (2.614,2.287)--(2.794,2.287);
\draw[gp path] (2.743,2.470)--(2.743,2.476);
\draw[gp path] (2.653,2.470)--(2.833,2.470);
\draw[gp path] (2.653,2.476)--(2.833,2.476);
\draw[gp path] (2.783,2.643)--(2.783,2.648);
\draw[gp path] (2.693,2.643)--(2.873,2.643);
\draw[gp path] (2.693,2.648)--(2.873,2.648);
\draw[gp path] (2.822,2.801)--(2.822,2.806);
\draw[gp path] (2.732,2.801)--(2.912,2.801);
\draw[gp path] (2.732,2.806)--(2.912,2.806);
\draw[gp path] (2.861,2.946)--(2.861,2.951);
\draw[gp path] (2.771,2.946)--(2.951,2.946);
\draw[gp path] (2.771,2.951)--(2.951,2.951);
\draw[gp path] (2.901,3.079)--(2.901,3.084);
\draw[gp path] (2.811,3.079)--(2.991,3.079);
\draw[gp path] (2.811,3.084)--(2.991,3.084);
\draw[gp path] (2.940,3.201)--(2.940,3.206);
\draw[gp path] (2.850,3.201)--(3.030,3.201);
\draw[gp path] (2.850,3.206)--(3.030,3.206);
\draw[gp path] (2.979,3.313)--(2.979,3.318);
\draw[gp path] (2.889,3.313)--(3.069,3.313);
\draw[gp path] (2.889,3.318)--(3.069,3.318);
\draw[gp path] (3.471,4.177)--(3.471,4.181);
\draw[gp path] (3.381,4.177)--(3.561,4.177);
\draw[gp path] (3.381,4.181)--(3.561,4.181);
\draw[gp path] (3.963,4.560)--(3.963,4.566);
\draw[gp path] (3.873,4.560)--(4.053,4.560);
\draw[gp path] (3.873,4.566)--(4.053,4.566);
\draw[gp path] (4.455,4.769)--(4.455,4.776);
\draw[gp path] (4.365,4.769)--(4.545,4.769);
\draw[gp path] (4.365,4.776)--(4.545,4.776);
\draw[gp path] (4.947,4.898)--(4.947,4.906);
\draw[gp path] (4.857,4.898)--(5.037,4.898);
\draw[gp path] (4.857,4.906)--(5.037,4.906);
\draw[gp path] (5.439,4.987)--(5.439,4.994);
\draw[gp path] (5.349,4.987)--(5.529,4.987);
\draw[gp path] (5.349,4.994)--(5.529,4.994);
\draw[gp path] (5.930,5.051)--(5.930,5.058);
\draw[gp path] (5.840,5.051)--(6.020,5.051);
\draw[gp path] (5.840,5.058)--(6.020,5.058);
\draw[gp path] (6.422,5.099)--(6.422,5.107);
\draw[gp path] (6.332,5.099)--(6.512,5.099);
\draw[gp path] (6.332,5.107)--(6.512,5.107);
\draw[gp path] (7.406,5.169)--(7.406,5.178);
\draw[gp path] (7.316,5.169)--(7.496,5.169);
\draw[gp path] (7.316,5.178)--(7.496,5.178);
\draw[gp path] (8.389,5.216)--(8.389,5.226);
\draw[gp path] (8.299,5.216)--(8.479,5.216);
\draw[gp path] (8.299,5.226)--(8.479,5.226);
\draw[gp path] (9.373,5.251)--(9.373,5.261);
\draw[gp path] (9.283,5.251)--(9.463,5.251);
\draw[gp path] (9.283,5.261)--(9.463,5.261);
\draw[gp path] (10.357,5.278)--(10.357,5.288);
\draw[gp path] (10.267,5.278)--(10.447,5.278);
\draw[gp path] (10.267,5.288)--(10.447,5.288);
\draw[gp path] (11.340,5.299)--(11.340,5.309);
\draw[gp path] (11.250,5.299)--(11.430,5.299);
\draw[gp path] (11.250,5.309)--(11.430,5.309);
\gppoint{gp mark 5}{(2.193,0.989)}
\gppoint{gp mark 5}{(2.232,0.991)}
\gppoint{gp mark 5}{(2.271,0.994)}
\gppoint{gp mark 5}{(2.311,0.997)}
\gppoint{gp mark 5}{(2.350,1.002)}
\gppoint{gp mark 5}{(2.389,1.008)}
\gppoint{gp mark 5}{(2.429,1.017)}
\gppoint{gp mark 5}{(2.468,1.029)}
\gppoint{gp mark 5}{(2.488,1.045)}
\gppoint{gp mark 5}{(2.507,1.138)}
\gppoint{gp mark 5}{(2.547,1.376)}
\gppoint{gp mark 5}{(2.586,1.622)}
\gppoint{gp mark 5}{(2.625,1.858)}
\gppoint{gp mark 5}{(2.665,2.079)}
\gppoint{gp mark 5}{(2.704,2.284)}
\gppoint{gp mark 5}{(2.743,2.473)}
\gppoint{gp mark 5}{(2.783,2.645)}
\gppoint{gp mark 5}{(2.822,2.804)}
\gppoint{gp mark 5}{(2.861,2.948)}
\gppoint{gp mark 5}{(2.901,3.081)}
\gppoint{gp mark 5}{(2.940,3.203)}
\gppoint{gp mark 5}{(2.979,3.315)}
\gppoint{gp mark 5}{(3.471,4.179)}
\gppoint{gp mark 5}{(3.963,4.563)}
\gppoint{gp mark 5}{(4.455,4.772)}
\gppoint{gp mark 5}{(4.947,4.902)}
\gppoint{gp mark 5}{(5.439,4.990)}
\gppoint{gp mark 5}{(5.930,5.055)}
\gppoint{gp mark 5}{(6.422,5.103)}
\gppoint{gp mark 5}{(7.406,5.173)}
\gppoint{gp mark 5}{(8.389,5.221)}
\gppoint{gp mark 5}{(9.373,5.256)}
\gppoint{gp mark 5}{(10.357,5.283)}
\gppoint{gp mark 5}{(11.340,5.304)}
\gppoint{gp mark 5}{(10.711,1.446)}
\draw[gp path] (2.193,0.989)--(2.287,0.995)--(2.381,1.007)--(2.476,1.029)--(2.570,1.523)%
  --(2.664,2.078)--(2.759,2.543)--(2.853,2.920)--(2.948,3.226)--(3.042,3.477)--(3.136,3.685)%
  --(3.231,3.858)--(3.325,4.001)--(3.420,4.121)--(3.514,4.224)--(3.608,4.314)--(3.703,4.392)%
  --(3.797,4.461)--(3.892,4.522)--(3.986,4.575)--(4.080,4.624)--(4.175,4.667)--(4.269,4.706)%
  --(4.363,4.741)--(4.458,4.773)--(4.552,4.803)--(4.647,4.830)--(4.741,4.854)--(4.835,4.877)%
  --(4.930,4.899)--(5.024,4.918)--(5.119,4.937)--(5.213,4.954)--(5.307,4.970)--(5.402,4.985)%
  --(5.496,4.999)--(5.591,5.012)--(5.685,5.025)--(5.779,5.037)--(5.874,5.048)--(5.968,5.059)%
  --(6.062,5.069)--(6.157,5.079)--(6.251,5.088)--(6.346,5.096)--(6.440,5.105)--(6.534,5.113)%
  --(6.629,5.120)--(6.723,5.128)--(6.818,5.135)--(6.912,5.142)--(7.006,5.148)--(7.101,5.155)%
  --(7.195,5.161)--(7.290,5.166)--(7.384,5.172)--(7.478,5.177)--(7.573,5.183)--(7.667,5.188)%
  --(7.761,5.192)--(7.856,5.197)--(7.950,5.202)--(8.045,5.206)--(8.139,5.210)--(8.233,5.214)%
  --(8.328,5.218)--(8.422,5.222)--(8.517,5.226)--(8.611,5.230)--(8.705,5.233)--(8.800,5.237)%
  --(8.894,5.240)--(8.989,5.243)--(9.083,5.246)--(9.177,5.250)--(9.272,5.253)--(9.366,5.256)%
  --(9.460,5.258)--(9.555,5.261)--(9.649,5.264)--(9.744,5.267)--(9.838,5.269)--(9.932,5.272)%
  --(10.027,5.274)--(10.121,5.277)--(10.216,5.279)--(10.310,5.282)--(10.404,5.284)--(10.499,5.286)%
  --(10.593,5.288)--(10.688,5.290)--(10.782,5.292)--(10.876,5.295)--(10.971,5.297)--(11.065,5.299)%
  --(11.159,5.300)--(11.254,5.302)--(11.348,5.304)--(11.443,5.306)--(11.537,5.308);
\gpcolor{color=gp lt color border}
\gpsetlinetype{gp lt border}
\gpsetlinewidth{1.00}
\draw[gp path] (1.504,6.895)--(1.504,0.985)--(11.537,0.985)--(11.537,6.895)--cycle;
\gpdefrectangularnode{gp plot 1}{\pgfpoint{1.504cm}{0.985cm}}{\pgfpoint{11.537cm}{6.895cm}}
\end{tikzpicture}

%% file: figures/G_k_kappa.tex
\begin{tikzpicture}[gnuplot]
{\scalefont{0.8}
\begin{scope}[scale=.8]
\gpsolidlines
\path (0.000,0.000) rectangle (12.090,7.264);
\gpcolor{color=gp lt color border}
\gpsetlinetype{gp lt border}
\gpsetlinewidth{1.00}
\draw[gp path] (1.688,0.985)--(1.868,0.985);
\draw[gp path] (11.537,0.985)--(11.357,0.985);
\node[gp node right] at (1.504,0.985) { 0.6};
\draw[gp path] (1.688,1.970)--(1.868,1.970);
\draw[gp path] (11.537,1.970)--(11.357,1.970);
\node[gp node right] at (1.504,1.970) { 0.65};
\draw[gp path] (1.688,2.955)--(1.868,2.955);
\draw[gp path] (11.537,2.955)--(11.357,2.955);
\node[gp node right] at (1.504,2.955) { 0.7};
\draw[gp path] (1.688,3.940)--(1.868,3.940);
\draw[gp path] (11.537,3.940)--(11.357,3.940);
\node[gp node right] at (1.504,3.940) { 0.75};
\draw[gp path] (1.688,4.925)--(1.868,4.925);
\draw[gp path] (11.537,4.925)--(11.357,4.925);
\node[gp node right] at (1.504,4.925) { 0.8};
\draw[gp path] (1.688,5.910)--(1.868,5.910);
\draw[gp path] (11.537,5.910)--(11.357,5.910);
\node[gp node right] at (1.504,5.910) { 0.85};
\draw[gp path] (1.688,6.895)--(1.868,6.895);
\draw[gp path] (11.537,6.895)--(11.357,6.895);
\node[gp node right] at (1.504,6.895) { 0.9};
\draw[gp path] (1.688,0.985)--(1.688,1.165);
\draw[gp path] (1.688,6.895)--(1.688,6.715);
\node[gp node center] at (1.688,0.677) { 0};
\draw[gp path] (3.095,0.985)--(3.095,1.165);
\draw[gp path] (3.095,6.895)--(3.095,6.715);
\node[gp node center] at (3.095,0.677) { 0.1};
\draw[gp path] (4.502,0.985)--(4.502,1.165);
\draw[gp path] (4.502,6.895)--(4.502,6.715);
\node[gp node center] at (4.502,0.677) { 0.2};
\draw[gp path] (5.909,0.985)--(5.909,1.165);
\draw[gp path] (5.909,6.895)--(5.909,6.715);
\node[gp node center] at (5.909,0.677) { 0.3};
\draw[gp path] (7.316,0.985)--(7.316,1.165);
\draw[gp path] (7.316,6.895)--(7.316,6.715);
\node[gp node center] at (7.316,0.677) { 0.4};
\draw[gp path] (8.723,0.985)--(8.723,1.165);
\draw[gp path] (8.723,6.895)--(8.723,6.715);
\node[gp node center] at (8.723,0.677) { 0.5};
\draw[gp path] (10.130,0.985)--(10.130,1.165);
\draw[gp path] (10.130,6.895)--(10.130,6.715);
\node[gp node center] at (10.130,0.677) { 0.6};
\draw[gp path] (11.537,0.985)--(11.537,1.165);
\draw[gp path] (11.537,6.895)--(11.537,6.715);
\node[gp node center] at (11.537,0.677) { 0.7};
\draw[gp path] (1.688,6.895)--(1.688,0.985)--(11.537,0.985)--(11.537,6.895)--cycle;
\node[gp node center,rotate=-270] at (0.246,3.940) {$G^\mathrm{E}/T^4$};
\node[gp node center] at (6.612,0.215) {$(q/T)^2$};
\node[gp node right] at (10.069,1.840) {$T=9.4T_\mathrm{c}$};
\gpcolor{rgb color={0.000,0.000,1.000}}
\gpsetlinetype{gp lt plot 0}
\gpsetlinewidth{3.00}
\draw[gp path] (10.253,1.840)--(11.169,1.840);
\draw[gp path] (10.253,1.930)--(10.253,1.750);
\draw[gp path] (11.169,1.930)--(11.169,1.750);
\draw[gp path] (2.035,1.330)--(2.035,3.749);
\draw[gp path] (1.945,1.330)--(2.125,1.330);
\draw[gp path] (1.945,3.749)--(2.125,3.749);
\draw[gp path] (3.077,2.147)--(3.077,4.567);
\draw[gp path] (2.987,2.147)--(3.167,2.147);
\draw[gp path] (2.987,4.567)--(3.167,4.567);
\draw[gp path] (4.812,3.197)--(4.812,5.613);
\draw[gp path] (4.722,3.197)--(4.902,3.197);
\draw[gp path] (4.722,5.613)--(4.902,5.613);
\draw[gp path] (7.243,2.776)--(7.243,5.175);
\draw[gp path] (7.153,2.776)--(7.333,2.776);
\draw[gp path] (7.153,5.175)--(7.333,5.175);
\draw[gp path] (10.367,4.113)--(10.367,6.521);
\draw[gp path] (10.277,4.113)--(10.457,4.113);
\draw[gp path] (10.277,6.521)--(10.457,6.521);
\gpsetpointsize{4.80}
\gppoint{gp mark 5}{(2.035,2.539)}
\gppoint{gp mark 5}{(3.077,3.357)}
\gppoint{gp mark 5}{(4.812,4.405)}
\gppoint{gp mark 5}{(7.243,3.975)}
\gppoint{gp mark 5}{(10.367,5.317)}
\gppoint{gp mark 5}{(10.711,1.840)}
\draw[gp path] (2.035,2.939)--(2.119,2.963)--(2.203,2.987)--(2.288,3.011)--(2.372,3.034)%
  --(2.456,3.058)--(2.540,3.082)--(2.624,3.105)--(2.708,3.129)--(2.793,3.153)--(2.877,3.177)%
  --(2.961,3.200)--(3.045,3.224)--(3.129,3.248)--(3.213,3.271)--(3.298,3.295)--(3.382,3.319)%
  --(3.466,3.342)--(3.550,3.366)--(3.634,3.390)--(3.718,3.414)--(3.803,3.437)--(3.887,3.461)%
  --(3.971,3.485)--(4.055,3.508)--(4.139,3.532)--(4.223,3.556)--(4.308,3.580)--(4.392,3.603)%
  --(4.476,3.627)--(4.560,3.651)--(4.644,3.674)--(4.728,3.698)--(4.812,3.722)--(4.897,3.746)%
  --(4.981,3.769)--(5.065,3.793)--(5.149,3.817)--(5.233,3.840)--(5.317,3.864)--(5.402,3.888)%
  --(5.486,3.912)--(5.570,3.935)--(5.654,3.959)--(5.738,3.983)--(5.822,4.006)--(5.907,4.030)%
  --(5.991,4.054)--(6.075,4.078)--(6.159,4.101)--(6.243,4.125)--(6.327,4.149)--(6.412,4.172)%
  --(6.496,4.196)--(6.580,4.220)--(6.664,4.244)--(6.748,4.267)--(6.832,4.291)--(6.916,4.315)%
  --(7.001,4.338)--(7.085,4.362)--(7.169,4.386)--(7.253,4.410)--(7.337,4.433)--(7.421,4.457)%
  --(7.506,4.481)--(7.590,4.504)--(7.674,4.528)--(7.758,4.552)--(7.842,4.576)--(7.926,4.599)%
  --(8.011,4.623)--(8.095,4.647)--(8.179,4.670)--(8.263,4.694)--(8.347,4.718)--(8.431,4.741)%
  --(8.516,4.765)--(8.600,4.789)--(8.684,4.813)--(8.768,4.836)--(8.852,4.860)--(8.936,4.884)%
  --(9.021,4.907)--(9.105,4.931)--(9.189,4.955)--(9.273,4.979)--(9.357,5.002)--(9.441,5.026)%
  --(9.525,5.050)--(9.610,5.073)--(9.694,5.097)--(9.778,5.121)--(9.862,5.145)--(9.946,5.168)%
  --(10.030,5.192)--(10.115,5.216)--(10.199,5.239)--(10.283,5.263)--(10.367,5.287);
\gpcolor{color=gp lt color border}
\node[gp node right] at (10.069,1.390) {LPT};
\gpcolor{rgb color={1.000,0.000,0.000}}
\gpsetlinetype{gp lt plot 2}
\draw[gp path] (10.253,1.390)--(11.169,1.390);
\draw[gp path] (2.035,1.915)--(2.119,1.942)--(2.203,1.970)--(2.288,1.998)--(2.372,2.026)%
  --(2.456,2.053)--(2.540,2.081)--(2.624,2.109)--(2.708,2.137)--(2.793,2.164)--(2.877,2.192)%
  --(2.961,2.220)--(3.045,2.248)--(3.129,2.275)--(3.213,2.303)--(3.298,2.331)--(3.382,2.358)%
  --(3.466,2.386)--(3.550,2.414)--(3.634,2.442)--(3.718,2.469)--(3.803,2.497)--(3.887,2.525)%
  --(3.971,2.553)--(4.055,2.580)--(4.139,2.608)--(4.223,2.636)--(4.308,2.664)--(4.392,2.691)%
  --(4.476,2.719)--(4.560,2.747)--(4.644,2.775)--(4.728,2.802)--(4.812,2.830)--(4.897,2.858)%
  --(4.981,2.886)--(5.065,2.913)--(5.149,2.941)--(5.233,2.969)--(5.317,2.997)--(5.402,3.024)%
  --(5.486,3.052)--(5.570,3.080)--(5.654,3.107)--(5.738,3.135)--(5.822,3.163)--(5.907,3.191)%
  --(5.991,3.218)--(6.075,3.246)--(6.159,3.274)--(6.243,3.302)--(6.327,3.329)--(6.412,3.357)%
  --(6.496,3.385)--(6.580,3.413)--(6.664,3.440)--(6.748,3.468)--(6.832,3.496)--(6.916,3.524)%
  --(7.001,3.551)--(7.085,3.579)--(7.169,3.607)--(7.253,3.635)--(7.337,3.662)--(7.421,3.690)%
  --(7.506,3.718)--(7.590,3.746)--(7.674,3.773)--(7.758,3.801)--(7.842,3.829)--(7.926,3.857)%
  --(8.011,3.884)--(8.095,3.912)--(8.179,3.940)--(8.263,3.967)--(8.347,3.995)--(8.431,4.023)%
  --(8.516,4.051)--(8.600,4.078)--(8.684,4.106)--(8.768,4.134)--(8.852,4.162)--(8.936,4.189)%
  --(9.021,4.217)--(9.105,4.245)--(9.189,4.273)--(9.273,4.300)--(9.357,4.328)--(9.441,4.356)%
  --(9.525,4.384)--(9.610,4.411)--(9.694,4.439)--(9.778,4.467)--(9.862,4.495)--(9.946,4.522)%
  --(10.030,4.550)--(10.115,4.578)--(10.199,4.606)--(10.283,4.633)--(10.367,4.661);
\gpcolor{color=gp lt color border}
\gpsetlinetype{gp lt border}
\gpsetlinewidth{1.00}
\draw[gp path] (1.688,6.895)--(1.688,0.985)--(11.537,0.985)--(11.537,6.895)--cycle;
\gpdefrectangularnode{gp plot 1}{\pgfpoint{1.688cm}{0.985cm}}{\pgfpoint{11.537cm}{6.895cm}}
\end{scope}}
\end{tikzpicture}

%% file: figures/T_kappa.tex
\begin{tikzpicture}[gnuplot]
\gpsolidlines
\path (0.000,0.000) rectangle (12.090,7.264);
\gpcolor{color=gp lt color border}
\gpsetlinetype{gp lt border}
\gpsetlinewidth{1.00}
\draw[gp path] (1.504,0.985)--(1.594,0.985);
\draw[gp path] (11.537,0.985)--(11.447,0.985);
\draw[gp path] (1.504,1.254)--(1.594,1.254);
\draw[gp path] (11.537,1.254)--(11.447,1.254);
\draw[gp path] (1.504,1.522)--(1.594,1.522);
\draw[gp path] (11.537,1.522)--(11.447,1.522);
\draw[gp path] (1.504,1.791)--(1.684,1.791);
\draw[gp path] (11.537,1.791)--(11.357,1.791);
\node[gp node right] at (1.320,1.791) {-0.5};
\draw[gp path] (1.504,2.060)--(1.594,2.060);
\draw[gp path] (11.537,2.060)--(11.447,2.060);
\draw[gp path] (1.504,2.328)--(1.594,2.328);
\draw[gp path] (11.537,2.328)--(11.447,2.328);
\draw[gp path] (1.504,2.597)--(1.594,2.597);
\draw[gp path] (11.537,2.597)--(11.447,2.597);
\draw[gp path] (1.504,2.865)--(1.594,2.865);
\draw[gp path] (11.537,2.865)--(11.447,2.865);
\draw[gp path] (1.504,3.134)--(1.684,3.134);
\draw[gp path] (11.537,3.134)--(11.357,3.134);
\node[gp node right] at (1.320,3.134) { 0};
\draw[gp path] (1.504,3.403)--(1.594,3.403);
\draw[gp path] (11.537,3.403)--(11.447,3.403);
\draw[gp path] (1.504,3.671)--(1.594,3.671);
\draw[gp path] (11.537,3.671)--(11.447,3.671);
\draw[gp path] (1.504,3.940)--(1.594,3.940);
\draw[gp path] (11.537,3.940)--(11.447,3.940);
\draw[gp path] (1.504,4.209)--(1.594,4.209);
\draw[gp path] (11.537,4.209)--(11.447,4.209);
\draw[gp path] (1.504,4.477)--(1.684,4.477);
\draw[gp path] (11.537,4.477)--(11.357,4.477);
\node[gp node right] at (1.320,4.477) { 0.5};
\draw[gp path] (1.504,4.746)--(1.594,4.746);
\draw[gp path] (11.537,4.746)--(11.447,4.746);
\draw[gp path] (1.504,5.015)--(1.594,5.015);
\draw[gp path] (11.537,5.015)--(11.447,5.015);
\draw[gp path] (1.504,5.283)--(1.594,5.283);
\draw[gp path] (11.537,5.283)--(11.447,5.283);
\draw[gp path] (1.504,5.552)--(1.594,5.552);
\draw[gp path] (11.537,5.552)--(11.447,5.552);
\draw[gp path] (1.504,5.820)--(1.684,5.820);
\draw[gp path] (11.537,5.820)--(11.357,5.820);
\node[gp node right] at (1.320,5.820) { 1};
\draw[gp path] (1.504,6.089)--(1.594,6.089);
\draw[gp path] (11.537,6.089)--(11.447,6.089);
\draw[gp path] (1.504,6.358)--(1.594,6.358);
\draw[gp path] (11.537,6.358)--(11.447,6.358);
\draw[gp path] (1.504,6.626)--(1.594,6.626);
\draw[gp path] (11.537,6.626)--(11.447,6.626);
\draw[gp path] (1.504,6.895)--(1.594,6.895);
\draw[gp path] (11.537,6.895)--(11.447,6.895);
\draw[gp path] (1.504,0.985)--(1.504,1.165);
\draw[gp path] (1.504,6.895)--(1.504,6.715);
\node[gp node center] at (1.504,0.677) { 1};
\draw[gp path] (1.705,0.985)--(1.705,1.075);
\draw[gp path] (1.705,6.895)--(1.705,6.805);
\draw[gp path] (1.905,0.985)--(1.905,1.075);
\draw[gp path] (1.905,6.895)--(1.905,6.805);
\draw[gp path] (2.106,0.985)--(2.106,1.075);
\draw[gp path] (2.106,6.895)--(2.106,6.805);
\draw[gp path] (2.307,0.985)--(2.307,1.075);
\draw[gp path] (2.307,6.895)--(2.307,6.805);
\draw[gp path] (2.507,0.985)--(2.507,1.165);
\draw[gp path] (2.507,6.895)--(2.507,6.715);
\node[gp node center] at (2.507,0.677) { 2};
\draw[gp path] (2.708,0.985)--(2.708,1.075);
\draw[gp path] (2.708,6.895)--(2.708,6.805);
\draw[gp path] (2.909,0.985)--(2.909,1.075);
\draw[gp path] (2.909,6.895)--(2.909,6.805);
\draw[gp path] (3.109,0.985)--(3.109,1.075);
\draw[gp path] (3.109,6.895)--(3.109,6.805);
\draw[gp path] (3.310,0.985)--(3.310,1.075);
\draw[gp path] (3.310,6.895)--(3.310,6.805);
\draw[gp path] (3.511,0.985)--(3.511,1.165);
\draw[gp path] (3.511,6.895)--(3.511,6.715);
\node[gp node center] at (3.511,0.677) { 3};
\draw[gp path] (3.711,0.985)--(3.711,1.075);
\draw[gp path] (3.711,6.895)--(3.711,6.805);
\draw[gp path] (3.912,0.985)--(3.912,1.075);
\draw[gp path] (3.912,6.895)--(3.912,6.805);
\draw[gp path] (4.113,0.985)--(4.113,1.075);
\draw[gp path] (4.113,6.895)--(4.113,6.805);
\draw[gp path] (4.313,0.985)--(4.313,1.075);
\draw[gp path] (4.313,6.895)--(4.313,6.805);
\draw[gp path] (4.514,0.985)--(4.514,1.165);
\draw[gp path] (4.514,6.895)--(4.514,6.715);
\node[gp node center] at (4.514,0.677) { 4};
\draw[gp path] (4.715,0.985)--(4.715,1.075);
\draw[gp path] (4.715,6.895)--(4.715,6.805);
\draw[gp path] (4.915,0.985)--(4.915,1.075);
\draw[gp path] (4.915,6.895)--(4.915,6.805);
\draw[gp path] (5.116,0.985)--(5.116,1.075);
\draw[gp path] (5.116,6.895)--(5.116,6.805);
\draw[gp path] (5.317,0.985)--(5.317,1.075);
\draw[gp path] (5.317,6.895)--(5.317,6.805);
\draw[gp path] (5.517,0.985)--(5.517,1.165);
\draw[gp path] (5.517,6.895)--(5.517,6.715);
\node[gp node center] at (5.517,0.677) { 5};
\draw[gp path] (5.718,0.985)--(5.718,1.075);
\draw[gp path] (5.718,6.895)--(5.718,6.805);
\draw[gp path] (5.919,0.985)--(5.919,1.075);
\draw[gp path] (5.919,6.895)--(5.919,6.805);
\draw[gp path] (6.119,0.985)--(6.119,1.075);
\draw[gp path] (6.119,6.895)--(6.119,6.805);
\draw[gp path] (6.320,0.985)--(6.320,1.075);
\draw[gp path] (6.320,6.895)--(6.320,6.805);
\draw[gp path] (6.521,0.985)--(6.521,1.165);
\draw[gp path] (6.521,6.895)--(6.521,6.715);
\node[gp node center] at (6.521,0.677) { 6};
\draw[gp path] (6.721,0.985)--(6.721,1.075);
\draw[gp path] (6.721,6.895)--(6.721,6.805);
\draw[gp path] (6.922,0.985)--(6.922,1.075);
\draw[gp path] (6.922,6.895)--(6.922,6.805);
\draw[gp path] (7.122,0.985)--(7.122,1.075);
\draw[gp path] (7.122,6.895)--(7.122,6.805);
\draw[gp path] (7.323,0.985)--(7.323,1.075);
\draw[gp path] (7.323,6.895)--(7.323,6.805);
\draw[gp path] (7.524,0.985)--(7.524,1.165);
\draw[gp path] (7.524,6.895)--(7.524,6.715);
\node[gp node center] at (7.524,0.677) { 7};
\draw[gp path] (7.724,0.985)--(7.724,1.075);
\draw[gp path] (7.724,6.895)--(7.724,6.805);
\draw[gp path] (7.925,0.985)--(7.925,1.075);
\draw[gp path] (7.925,6.895)--(7.925,6.805);
\draw[gp path] (8.126,0.985)--(8.126,1.075);
\draw[gp path] (8.126,6.895)--(8.126,6.805);
\draw[gp path] (8.326,0.985)--(8.326,1.075);
\draw[gp path] (8.326,6.895)--(8.326,6.805);
\draw[gp path] (8.527,0.985)--(8.527,1.165);
\draw[gp path] (8.527,6.895)--(8.527,6.715);
\node[gp node center] at (8.527,0.677) { 8};
\draw[gp path] (8.728,0.985)--(8.728,1.075);
\draw[gp path] (8.728,6.895)--(8.728,6.805);
\draw[gp path] (8.928,0.985)--(8.928,1.075);
\draw[gp path] (8.928,6.895)--(8.928,6.805);
\draw[gp path] (9.129,0.985)--(9.129,1.075);
\draw[gp path] (9.129,6.895)--(9.129,6.805);
\draw[gp path] (9.330,0.985)--(9.330,1.075);
\draw[gp path] (9.330,6.895)--(9.330,6.805);
\draw[gp path] (9.530,0.985)--(9.530,1.165);
\draw[gp path] (9.530,6.895)--(9.530,6.715);
\node[gp node center] at (9.530,0.677) { 9};
\draw[gp path] (9.731,0.985)--(9.731,1.075);
\draw[gp path] (9.731,6.895)--(9.731,6.805);
\draw[gp path] (9.932,0.985)--(9.932,1.075);
\draw[gp path] (9.932,6.895)--(9.932,6.805);
\draw[gp path] (10.132,0.985)--(10.132,1.075);
\draw[gp path] (10.132,6.895)--(10.132,6.805);
\draw[gp path] (10.333,0.985)--(10.333,1.075);
\draw[gp path] (10.333,6.895)--(10.333,6.805);
\draw[gp path] (10.534,0.985)--(10.534,1.165);
\draw[gp path] (10.534,6.895)--(10.534,6.715);
\node[gp node center] at (10.534,0.677) { 10};
\draw[gp path] (10.734,0.985)--(10.734,1.075);
\draw[gp path] (10.734,6.895)--(10.734,6.805);
\draw[gp path] (10.935,0.985)--(10.935,1.075);
\draw[gp path] (10.935,6.895)--(10.935,6.805);
\draw[gp path] (11.136,0.985)--(11.136,1.075);
\draw[gp path] (11.136,6.895)--(11.136,6.805);
\draw[gp path] (11.336,0.985)--(11.336,1.075);
\draw[gp path] (11.336,6.895)--(11.336,6.805);
\draw[gp path] (11.537,0.985)--(11.537,1.165);
\draw[gp path] (11.537,6.895)--(11.537,6.715);
\node[gp node center] at (11.537,0.677) { 11};
\draw[gp path] (1.504,6.895)--(1.504,0.985)--(11.537,0.985)--(11.537,6.895)--cycle;
\node[gp node center,rotate=-270] at (0.246,3.940) {$\kappa/T^2$};
\node[gp node center] at (6.520,0.215) {$T/T_\mathrm{c}$};
\node[gp node right] at (10.069,3.190) {$\beta=7.1$};
\gpcolor{rgb color={1.000,0.000,0.000}}
\gpsetlinetype{gp lt plot 0}
\gpsetlinewidth{3.00}
\draw[gp path] (10.253,3.190)--(11.169,3.190);
\draw[gp path] (10.253,3.280)--(10.253,3.100);
\draw[gp path] (11.169,3.280)--(11.169,3.100);
\draw[gp path] (9.932,3.526)--(9.932,4.905);
\draw[gp path] (9.842,3.526)--(10.022,3.526);
\draw[gp path] (9.842,4.905)--(10.022,4.905);
\draw[gp path] (7.624,1.974)--(7.624,6.476);
\draw[gp path] (7.534,1.974)--(7.714,1.974);
\draw[gp path] (7.534,6.476)--(7.714,6.476);
\gpsetpointsize{4.80}
\gppoint{gp mark 5}{(9.932,4.215)}
\gppoint{gp mark 5}{(7.624,4.225)}
\gppoint{gp mark 5}{(10.711,3.190)}
\gpcolor{color=gp lt color border}
\node[gp node right] at (10.069,2.740) {$\beta=6.68$};
\gpcolor{rgb color={0.000,0.000,1.000}}
\gpsetlinetype{gp lt plot 1}
\draw[gp path] (10.253,2.740)--(11.169,2.740);
\draw[gp path] (10.253,2.830)--(10.253,2.650);
\draw[gp path] (11.169,2.830)--(11.169,2.650);
\draw[gp path] (6.119,3.390)--(6.119,4.986);
\draw[gp path] (6.029,3.390)--(6.209,3.390);
\draw[gp path] (6.029,4.986)--(6.209,4.986);
\gppoint{gp mark 7}{(6.119,4.188)}
\gppoint{gp mark 7}{(10.711,2.740)}
\gpcolor{color=gp lt color border}
\node[gp node right] at (10.069,2.290) {$\beta=6.14$};
\gpcolor{rgb color={0.000,0.502,0.000}}
\gpsetlinetype{gp lt plot 3}
\draw[gp path] (10.253,2.290)--(11.169,2.290);
\draw[gp path] (10.253,2.380)--(10.253,2.200);
\draw[gp path] (11.169,2.380)--(11.169,2.200);
\draw[gp path] (3.109,3.348)--(3.109,4.426);
\draw[gp path] (3.019,3.348)--(3.199,3.348);
\draw[gp path] (3.019,4.426)--(3.199,4.426);
\gpsetpointsize{6.00}
\gppoint{gp mark 8}{(3.109,3.887)}
\gppoint{gp mark 8}{(10.711,2.290)}
\gpcolor{color=gp lt color border}
\node[gp node right] at (10.069,1.840) {ADS/CFT};
\gpcolor{rgb color={0.580,0.000,0.827}}
\gpsetlinetype{gp lt plot 2}
\draw[gp path] (10.253,1.840)--(11.169,1.840);
\draw[gp path] (10.253,1.930)--(10.253,1.750);
\draw[gp path] (11.169,1.930)--(11.169,1.750);
\draw[gp path] (1.504,3.208)--(1.504,3.209);
\draw[gp path] (1.414,3.208)--(1.594,3.208);
\draw[gp path] (1.414,3.209)--(1.594,3.209);
\draw[gp path] (1.434,3.291)--(1.614,3.291);
\draw[gp path] (1.434,3.291)--(1.614,3.291);
\draw[gp path] (1.474,3.339)--(1.654,3.339);
\draw[gp path] (1.474,3.339)--(1.654,3.339);
\draw[gp path] (1.604,3.367)--(1.604,3.368);
\draw[gp path] (1.514,3.367)--(1.694,3.367);
\draw[gp path] (1.514,3.368)--(1.694,3.368);
\draw[gp path] (1.554,3.388)--(1.734,3.388);
\draw[gp path] (1.554,3.388)--(1.734,3.388);
\draw[gp path] (1.685,3.404)--(1.685,3.405);
\draw[gp path] (1.595,3.404)--(1.775,3.404);
\draw[gp path] (1.595,3.405)--(1.775,3.405);
\draw[gp path] (1.725,3.417)--(1.725,3.418);
\draw[gp path] (1.635,3.417)--(1.815,3.417);
\draw[gp path] (1.635,3.418)--(1.815,3.418);
\draw[gp path] (1.765,3.428)--(1.765,3.429);
\draw[gp path] (1.675,3.428)--(1.855,3.428);
\draw[gp path] (1.675,3.429)--(1.855,3.429);
\draw[gp path] (1.805,3.438)--(1.805,3.439);
\draw[gp path] (1.715,3.438)--(1.895,3.438);
\draw[gp path] (1.715,3.439)--(1.895,3.439);
\draw[gp path] (1.845,3.446)--(1.845,3.447);
\draw[gp path] (1.755,3.446)--(1.935,3.446);
\draw[gp path] (1.755,3.447)--(1.935,3.447);
\draw[gp path] (1.795,3.454)--(1.975,3.454);
\draw[gp path] (1.795,3.454)--(1.975,3.454);
\draw[gp path] (1.835,3.461)--(2.015,3.461);
\draw[gp path] (1.835,3.461)--(2.015,3.461);
\draw[gp path] (1.876,3.467)--(2.056,3.467);
\draw[gp path] (1.876,3.467)--(2.056,3.467);
\draw[gp path] (2.006,3.472)--(2.006,3.473);
\draw[gp path] (1.916,3.472)--(2.096,3.472);
\draw[gp path] (1.916,3.473)--(2.096,3.473);
\draw[gp path] (2.507,3.512)--(2.507,3.513);
\draw[gp path] (2.417,3.512)--(2.597,3.512);
\draw[gp path] (2.417,3.513)--(2.597,3.513);
\draw[gp path] (2.919,3.531)--(3.099,3.531);
\draw[gp path] (2.919,3.531)--(3.099,3.531);
\draw[gp path] (3.511,3.542)--(3.511,3.543);
\draw[gp path] (3.421,3.542)--(3.601,3.542);
\draw[gp path] (3.421,3.543)--(3.601,3.543);
\draw[gp path] (4.012,3.549)--(4.012,3.550);
\draw[gp path] (3.922,3.549)--(4.102,3.549);
\draw[gp path] (3.922,3.550)--(4.102,3.550);
\draw[gp path] (4.424,3.555)--(4.604,3.555);
\draw[gp path] (4.424,3.555)--(4.604,3.555);
\draw[gp path] (4.926,3.559)--(5.106,3.559);
\draw[gp path] (4.926,3.559)--(5.106,3.559);
\draw[gp path] (5.517,3.562)--(5.517,3.563);
\draw[gp path] (5.427,3.562)--(5.607,3.562);
\draw[gp path] (5.427,3.563)--(5.607,3.563);
\draw[gp path] (6.521,3.567)--(6.521,3.568);
\draw[gp path] (6.431,3.567)--(6.611,3.567);
\draw[gp path] (6.431,3.568)--(6.611,3.568);
\draw[gp path] (7.524,3.570)--(7.524,3.571);
\draw[gp path] (7.434,3.570)--(7.614,3.570);
\draw[gp path] (7.434,3.571)--(7.614,3.571);
\draw[gp path] (8.527,3.572)--(8.527,3.573);
\draw[gp path] (8.437,3.572)--(8.617,3.572);
\draw[gp path] (8.437,3.573)--(8.617,3.573);
\draw[gp path] (9.530,3.574)--(9.530,3.575);
\draw[gp path] (9.440,3.574)--(9.620,3.574);
\draw[gp path] (9.440,3.575)--(9.620,3.575);
\draw[gp path] (10.534,3.576)--(10.534,3.577);
\draw[gp path] (10.444,3.576)--(10.624,3.576);
\draw[gp path] (10.444,3.577)--(10.624,3.577);
\gpsetpointsize{4.00}
\gppoint{gp mark 3}{(1.504,3.208)}
\gppoint{gp mark 3}{(1.524,3.291)}
\gppoint{gp mark 3}{(1.564,3.339)}
\gppoint{gp mark 3}{(1.604,3.368)}
\gppoint{gp mark 3}{(1.644,3.388)}
\gppoint{gp mark 3}{(1.685,3.404)}
\gppoint{gp mark 3}{(1.725,3.418)}
\gppoint{gp mark 3}{(1.765,3.429)}
\gppoint{gp mark 3}{(1.805,3.438)}
\gppoint{gp mark 3}{(1.845,3.447)}
\gppoint{gp mark 3}{(1.885,3.454)}
\gppoint{gp mark 3}{(1.925,3.461)}
\gppoint{gp mark 3}{(1.966,3.467)}
\gppoint{gp mark 3}{(2.006,3.472)}
\gppoint{gp mark 3}{(2.507,3.512)}
\gppoint{gp mark 3}{(3.009,3.531)}
\gppoint{gp mark 3}{(3.511,3.542)}
\gppoint{gp mark 3}{(4.012,3.550)}
\gppoint{gp mark 3}{(4.514,3.555)}
\gppoint{gp mark 3}{(5.016,3.559)}
\gppoint{gp mark 3}{(5.517,3.562)}
\gppoint{gp mark 3}{(6.521,3.567)}
\gppoint{gp mark 3}{(7.524,3.570)}
\gppoint{gp mark 3}{(8.527,3.573)}
\gppoint{gp mark 3}{(9.530,3.575)}
\gppoint{gp mark 3}{(10.534,3.577)}
\gppoint{gp mark 3}{(10.711,1.840)}
\draw[gp path] (1.504,3.208)--(1.605,3.368)--(1.707,3.412)--(1.808,3.439)--(1.909,3.458)%
  --(2.011,3.473)--(2.112,3.484)--(2.213,3.494)--(2.315,3.501)--(2.416,3.508)--(2.517,3.513)%
  --(2.619,3.518)--(2.720,3.522)--(2.821,3.525)--(2.923,3.529)--(3.024,3.531)--(3.125,3.534)%
  --(3.227,3.537)--(3.328,3.539)--(3.430,3.541)--(3.531,3.543)--(3.632,3.544)--(3.734,3.546)%
  --(3.835,3.547)--(3.936,3.549)--(4.038,3.550)--(4.139,3.551)--(4.240,3.552)--(4.342,3.553)%
  --(4.443,3.554)--(4.544,3.555)--(4.646,3.556)--(4.747,3.557)--(4.848,3.558)--(4.950,3.559)%
  --(5.051,3.559)--(5.152,3.560)--(5.254,3.561)--(5.355,3.561)--(5.456,3.562)--(5.558,3.563)%
  --(5.659,3.563)--(5.760,3.564)--(5.862,3.564)--(5.963,3.565)--(6.064,3.565)--(6.166,3.566)%
  --(6.267,3.566)--(6.368,3.566)--(6.470,3.567)--(6.571,3.567)--(6.673,3.568)--(6.774,3.568)%
  --(6.875,3.568)--(6.977,3.569)--(7.078,3.569)--(7.179,3.569)--(7.281,3.570)--(7.382,3.570)%
  --(7.483,3.570)--(7.585,3.571)--(7.686,3.571)--(7.787,3.571)--(7.889,3.571)--(7.990,3.572)%
  --(8.091,3.572)--(8.193,3.572)--(8.294,3.572)--(8.395,3.573)--(8.497,3.573)--(8.598,3.573)%
  --(8.699,3.573)--(8.801,3.574)--(8.902,3.574)--(9.003,3.574)--(9.105,3.574)--(9.206,3.574)%
  --(9.307,3.575)--(9.409,3.575)--(9.510,3.575)--(9.611,3.575)--(9.713,3.575)--(9.814,3.575)%
  --(9.916,3.576)--(10.017,3.576)--(10.118,3.576)--(10.220,3.576)--(10.321,3.576)--(10.422,3.576)%
  --(10.524,3.577)--(10.625,3.577)--(10.726,3.577)--(10.828,3.577)--(10.929,3.577)--(11.030,3.577)%
  --(11.132,3.577)--(11.233,3.578)--(11.334,3.578)--(11.436,3.578)--(11.537,3.578);
\gpcolor{color=gp lt color border}
\node[gp node right] at (10.069,1.390) {LPT};
\gpcolor{rgb color={0.502,0.502,0.502}}
\gpsetlinetype{gp lt axes}
\draw[gp path] (10.253,1.390)--(11.169,1.390);
\draw[gp path] (1.504,4.399)--(1.605,4.399)--(1.707,4.399)--(1.808,4.399)--(1.909,4.399)%
  --(2.011,4.399)--(2.112,4.399)--(2.213,4.399)--(2.315,4.399)--(2.416,4.399)--(2.517,4.399)%
  --(2.619,4.399)--(2.720,4.399)--(2.821,4.399)--(2.923,4.399)--(3.024,4.399)--(3.125,4.399)%
  --(3.227,4.399)--(3.328,4.399)--(3.430,4.399)--(3.531,4.399)--(3.632,4.399)--(3.734,4.399)%
  --(3.835,4.399)--(3.936,4.399)--(4.038,4.399)--(4.139,4.399)--(4.240,4.399)--(4.342,4.399)%
  --(4.443,4.399)--(4.544,4.399)--(4.646,4.399)--(4.747,4.399)--(4.848,4.399)--(4.950,4.399)%
  --(5.051,4.399)--(5.152,4.399)--(5.254,4.399)--(5.355,4.399)--(5.456,4.399)--(5.558,4.399)%
  --(5.659,4.399)--(5.760,4.399)--(5.862,4.399)--(5.963,4.399)--(6.064,4.399)--(6.166,4.399)%
  --(6.267,4.399)--(6.368,4.399)--(6.470,4.399)--(6.571,4.399)--(6.673,4.399)--(6.774,4.399)%
  --(6.875,4.399)--(6.977,4.399)--(7.078,4.399)--(7.179,4.399)--(7.281,4.399)--(7.382,4.399)%
  --(7.483,4.399)--(7.585,4.399)--(7.686,4.399)--(7.787,4.399)--(7.889,4.399)--(7.990,4.399)%
  --(8.091,4.399)--(8.193,4.399)--(8.294,4.399)--(8.395,4.399)--(8.497,4.399)--(8.598,4.399)%
  --(8.699,4.399)--(8.801,4.399)--(8.902,4.399)--(9.003,4.399)--(9.105,4.399)--(9.206,4.399)%
  --(9.307,4.399)--(9.409,4.399)--(9.510,4.399)--(9.611,4.399)--(9.713,4.399)--(9.814,4.399)%
  --(9.916,4.399)--(10.017,4.399)--(10.118,4.399)--(10.220,4.399)--(10.321,4.399)--(10.422,4.399)%
  --(10.524,4.399)--(10.625,4.399)--(10.726,4.399)--(10.828,4.399)--(10.929,4.399)--(11.030,4.399)%
  --(11.132,4.399)--(11.233,4.399)--(11.334,4.399)--(11.436,4.399)--(11.537,4.399);
\gpcolor{color=gp lt color border}
\gpsetlinetype{gp lt border}
\gpsetlinewidth{1.00}
\draw[gp path] (1.504,6.895)--(1.504,0.985)--(11.537,0.985)--(11.537,6.895)--cycle;
\node[gp node right] at (11.5,4.9) {$N_\tau=6$};
\node[gp node right] at (9.15,6.47) {$N_\tau=8$};
\node[gp node right] at (7.65,5.0) {$N_\tau=6$};
\node[gp node right] at (4.65,4.75) {$N_\tau=6$};
\gpdefrectangularnode{gp plot 1}{\pgfpoint{1.504cm}{0.985cm}}{\pgfpoint{11.537cm}{6.895cm}}
\end{tikzpicture}